\documentclass[journal]{IEEEtran}

\usepackage{subfigure}
\usepackage{amsmath,amssymb}
\usepackage[dvips]{graphicx}
\usepackage{amsfonts}
\usepackage[mathscr]{eucal}
\usepackage{latexsym}
\usepackage{amsthm}
\usepackage{exscale}
\usepackage[mathscr]{eucal}
\usepackage{bm}
\usepackage[dvipsnames]{color}
\usepackage{cases}
\usepackage{epsfig}
\usepackage[center,small]{caption}
\usepackage{algorithm}
\usepackage{algorithmic}
\usepackage[verbose,nospace,sort]{cite}
\usepackage{tabularx}
\usepackage{multirow}
\usepackage{multicol}
\usepackage{balance}

\graphicspath{{./figsJournal/}}

\scrollmode

\newtheorem{theorem}{Theorem}

\newtheorem{lemma}{Lemma}

\newcommand{\hSDsq}{|h_{SD}|^2}

\newcommand{\gammatd}{\tilde{\gamma}}
\newcommand{\lk}{\mathrm{ev}}
\newcommand{\Ptd}{\tilde{P}}
\newcommand{\diag}{\mathrm{diag}}
\newcommand{\xE}{\mathbb{E}}
\newcommand{\tdh}{\tilde{\mathbf h}}
\newcommand{\hath}{\hat{\mathbf h}}

\begin{document}
\title{Wireless Information Surveillance via Proactive Eavesdropping with Spoofing Relay}
\author{Yong~Zeng$^\ast$ and Rui~Zhang
\thanks{Y. Zeng (corresponding author) is with the Department of Electrical and Computer Engineering, National University of Singapore (e-mail: elezeng@nus.edu.sg).}
\thanks{R. Zhang is with the Department of Electrical and Computer Engineering, National University of Singapore (e-mail: elezhang@nus.edu.sg). He is also with the Institute for Infocomm Research, A*STAR, Singapore.}
\thanks{This work has been presented in part at the 41th IEEE International Conference on Acoustics, Speech and Signal Processing (ICASSP), 20-25 March 2016, Shanghai, China.}
}
\maketitle

\begin{abstract}
   Wireless information surveillance, by which suspicious wireless communications are closely monitored by legitimate agencies,  is an integral part of national security. To enhance the information surveillance capability, we propose in this paper a new proactive eavesdropping approach via a spoofing relay, where the legitimate monitor operates in a full-duplex manner with simultaneous eavesdropping and spoofing relaying to vary the source transmission rate in favor of the eavesdropping performance. To this end, a power splitting receiver is proposed, where the signal received at each antenna of the legitimate monitor is split into two parts for information eavesdropping and spoofing relaying, respectively. We formulate an optimization problem to maximize the achievable eavesdropping rate by jointly optimizing the power splitting ratios and  relay beamforming matrix at the multi-antenna   monitor.  Depending on the suspicious  and  eavesdropping channel conditions, the optimal solution corresponds to three possible spoofing relay strategies, namely  \emph{constructive relaying}, \emph{jamming}, and \emph{simultaneous jamming and destructive relaying}. Numerical results show that the proposed technique significantly improves the eavesdropping rate of the legitimate monitor as compared to the existing passive eavesdropping and jamming-based eavesdropping schemes.
\end{abstract}


\section{Introduction}
The great success of wireless communication technology has drastically improved our life during the past few decades. By providing high-speed wireless connectivity essentially anywhere, anytime, and between any pair of devices, contemporary wireless communication systems not only make our daily life increasingly more convenient, but also provide numerous new opportunities for applications and innovations in almost all fields, such as education, business, industry, etc. However, the ubiquitous accessibility of wireless communication systems also makes them more vulnerable to be misused by malicious users to commit crimes, jeopardize the public safety, and invade the privacy of others, etc. Therefore, it becomes increasingly important for the legitimate parties, such as the government agencies, to implement effective information surveillance measures to monitor any suspicious communication for various purposes such as intelligence gathering,  terrorism/crime prevention and investigation, etc.

From an engineering design perspective, devising efficient schemes for wireless information surveillance  calls for a paradigm shift from that for conventional wireless security \cite{648},\cite{644}. In wireless security, the eavesdroppers are treated as adversaries, whose eavesdropping potential should be minimized. Various wireless security mechanisms, both in physical layer \cite{596} and across upper layers of communication system protocols design~\cite{603}, have been proposed to prevent or minimize the information leakage to the unintended eavesdroppers. In particular, under the classic wiretap channel setup~\cite{599}, significant efforts have been devoted to characterizing the {\it secrecy capacity} \cite{601,607,602}, defined as the maximum transmission rate at which the message can be reliably decoded at the legitimate receiver without leaking any useful information to eavesdropping receivers. For wireless surveillance, however, the monitor is regarded as a legitimate eavesdropper, whose eavesdropping rate over any suspicious channel should be maximized to effectively intercept the transmitted  information over the air.

One straightforward approach for wireless information surveillance is passive eavesdropping, where the legitimate monitor only listens to the wireless channels of any suspicious users to decode their transmitted messages. However, this approach is effective only when the eavesdropping channel from the source to the legitimate monitor is better than the suspicious channel to the destination, so that the information sent by the suspicious source can be reliably decoded at the legitimate monitor. Yet this does not hold in practice if the legitimate monitor is located further away from the suspicious source compared to the suspicious destination. To overcome this limitation, a proactive eavesdropping scheme via cognitive jamming technique  is proposed in \cite{648}, \cite{644}, where the legitimate monitor sends noise-like jamming signals to intentionally degrade the suspicious link, so as to induce the suspicious source to reduce transmission rate to be decodable at the legitimate monitor receiver. However, given the limited power budget for jamming, the eavesdropping performance cannot be improved if the suspicious channel capacity is higher than that of the eavesdropping channel even with the maximum-power jamming. {  Furthermore, there are also scenarios when it is desirable for the legitimate monitor to spoof the suspicious source to increase its transmission rate to achieve higher eavesdropping rate. In such cases, jamming is not effective and more intelligent strategies need to be devised for proactive eavesdropping.}

\begin{figure}
\centering
\includegraphics[scale=0.7]{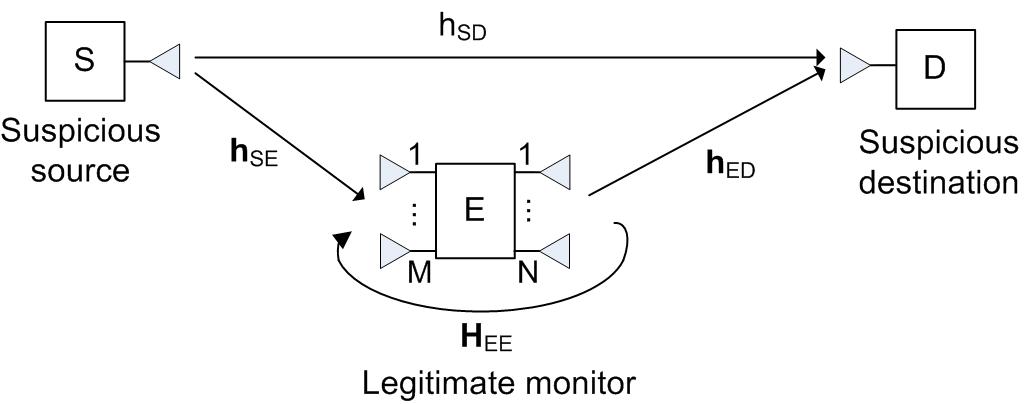}
\caption{Wireless information surveillance via a multi-antenna legitimate monitor.}\label{F:model}
\end{figure}

To further enhance the information surveillance capability of legitimate monitors, we propose in this paper a new proactive eavesdropping approach via a so-called \emph{spoofing relay} technique. Specifically, besides information eavesdropping, the legitimate monitor also acts concurrently as a relay  to send spoofing signals to the suspicious destination and thereby  induce the source to vary the transmission rate in favor of the eavesdropping performance. The underlying assumption is that adaptive rate transmission is adopted at the suspicious source based on the effective channel condition at the destination. Under this setup, if the eavesdropping link (from the source to the legitimate monitor) is stronger than the suspicious link (from the source to the destination), the legitimate monitor will enhance the effective channel of the suspicious link by forwarding a constructive source signal to the suspicious destination, which leads to higher transmission rate by the suspicious source, and thus higher eavesdropping rate. On the other hand, if the eavesdropping link  is weaker than the suspicious link, the legitimate monitor will degrade the effective channel of the suspicious link via forwarding a destructive source signal and/or a noise-like jamming signal to the suspicious destination, so as to spoof the suspicious source to reduce its transmission rate to a level decodable by the eavesdropping monitor. {  Note that the proposed spoofing relay technique is quite general since it is applicable regardless of whether the eavesdropping link is stronger or weaker than the suspicious link. Furthermore, it includes the existing passive and jamming-based eavesdropping \cite{648},\cite{644} as special cases, and thus is expected to outperform these two benchmark schemes. Intuitively, the proposed spoofing relay scheme is strictly beneficial when either the eavesdropping link is stronger than the suspicious link (so that constructive relaying improves the eavesdropping rate), or when the eavesdropping link is too weak such that jamming alone is insufficient (so destructive relaying is needed), as will be verified later in this paper by numerical results.} The main contributions of this paper are summarized as follows.

\begin{itemize}
\item First, we model the system architecture for wireless information surveillance via a legitimate multi-antenna monitor, as shown in Fig.~\ref{F:model}. A power splitting receiver\footnote{Notice that power splitting technique has also been used for separating the received signal for information decoding and energy harvesting in simultaneous wireless information and power transfer (SWIPT) systems \cite{478},\cite{514}.}  is proposed to split the received signal at each receiving antenna of the legitimate monitor into two parts, one for information eavesdropping and the other for spoofing relaying. An optimization problem is then formulated to maximize the eavesdropping rate by the legitimate monitor via jointly optimizing the power splitting ratios for all of its receiving antennas and the relay beamforming/precoding matrix for the transmitting antennas.
\item Next, we derive the optimal solution to the formulated problem by first solving two key sub-problems, which respectively find the maximum and minimum effective signal-to-noise ratio (SNR) at the suspicious link receiver by optimizing the relay precoding matrix at the legitimate monitor transmitter with fixed receiver power splitting ratios.
    Based on the obtained solutions, we further show that uniform power splitting, i.e., all receiving antennas at the legitimate monitor use the same power splitting ratio, is optimal for maximizing the eavesdropping rate with spoofing relaying, which thus leads to an efficient solution for the optimal power splitting ratios. Finally, the problem is solved with three possible relay  strategies at the legitimate monitor, namely  \emph{constructive relaying}, \emph{jamming}, and \emph{simultaneous jamming and destructive relaying}, which  are applied when the eavesdropping channel is better, weaker, and severely weaker than the suspicious channel, respectively.
\item 
    At last, numerical results are provided, which show that the  eavesdropping rate achievable by the proposed spoofing relay scheme is significantly higher than that of the  two benchmark schemes, namely passive eavesdropping and jamming-based proactive eavesdropping \cite{648},\cite{644}.
\end{itemize}


{  It is worth pointing out that under the classic physical-layer security framework, secure communication for wireless relay channels has been studied in various setups. Depending on the role of the relay node, such existing works can be loosely classified into three categories: (i) {\it trusted relay} that helps the source transmitter in improving the secrecy rate in the presence of the eavesdropper, via cooperative signal relaying to the destination or cooperative jamming to the eavesdropper \cite{769,770,771}; (ii) {\it untrusted relay} from which the source transmitter wishes to keep the information confidential while engaging its help \cite{772},\cite{773}; and (iii) {\it adversary relay} that helps the eavesdropper, rather than the source transmitter, in decreasing the secrecy rate \cite{774},\cite{775}. On the other hand, proactive eavesdropping in different forms such as pilot contamination attack, false feedback, etc., has also been studied recently \cite{646,604,647,608,605,609,606,610,611}. However, all the aforementioned works focus on the conventional wireless security design based on the information-theoretic secrecy capacity measure, which treats the eavesdroppers as adversaries and thus aims to minimize the information leakage to them.  In contrast, for the  proactive eavesdropping considered in this paper, the eavesdropper is employed by a legitimate monitor for the different purpose of information surveillance, and how to maximize its eavesdropping rate via optimally designing the spoofing relay strategy is a new problem that has not been studied in the literature.}

The rest of this paper is organized as follows. Section~\ref{sec:systemModel} introduces the system model for proactive eavesdropping with a spoofing relay, and presents the problem formulation for eavesdropping rate maximization. Section~\ref{eq:optSol} presents the optimal solution for the formulated problem, as well as a low-cost implementation that only requires one power splitter at the legitimate monitor receiver.  In Section~\ref{sec:numerical}, numerical results are presented to compare the proposed proactive eavesdropping scheme with the two benchmark schemes. Finally, we conclude the paper in Section~\ref{sec:conclusion}.

\emph{Notations:} In this paper, scalars are denoted by italic letters. Boldface lower- and upper-case letters denote vectors and matrices, respectively. $\mathbb{C}^{M\times N}$ denotes the space of $M\times N$ complex-valued matrices.   $\mathbf{I}$ represents an identity matrix, $\mathbf{0}$ and $\mathbf 1$ denote an all-zero and all-one matrix, respectively. For a matrix $\mathbf{A}$,  its complex conjugate, transpose, Hermitian transpose, and Frobenius norm are respectively denoted as $\mathbf A^*$, $\mathbf{A}^{T}$, $\mathbf{A}^{H}$, and $\|\mathbf A\|_F$. For a vector $\mathbf a$, $\|\mathbf a\|$ represents its Euclidean norm. 
  $\diag{(\mathbf a)}$ denotes a diagonal matrix with the diagonal elements given in the vector $\mathbf a$. $\mathbf a \preceq \mathbf b$ means that each element in $\mathbf a$ is no greater than that in $\mathbf b$. For a complex number $z$, $\angle z$ represents its phase, and $\Re(z)$ and $\Im(z)$ denote its real and imaginary parts, respectively.  The symbol $j$ represents the imaginary unit of complex numbers, i.e., $j^2=-1$.

\section{System Model and Problem Formulation}\label{sec:systemModel}
As shown in Fig.~\ref{F:model}, we consider an information surveillance scenario, where the legitimate monitor {\bf E} is intended to overhear a suspicious communication link from source {\bf S} to destination {\bf D}. We assume that both {\bf S} and {\bf D} have one single antenna each, whereas {\bf E} is equipped with $M\geq 1$ receiving and $N\geq 1$ transmitting antennas. We consider that {\it adaptive rate transmission} is adopted at {\bf S} based on the channel perceived at {\bf D}. However, neither {\bf S} nor {\bf D} is aware of the presence of {\bf E}, so that no dedicated coding as in conventional physical-layer security  (see e.g., \cite{596} and references therein) is applied to prevent the eavesdropping by {\bf E}. On the other hand, the legitimate monitor {\bf E} can conduct either passive or proactive eavesdropping, as discussed below.

\subsection{Passive Eavesdropping}\label{sec:passive}
With passive eavesdropping, {\bf E} remains silent throughout the communication between {\bf S} and {\bf D}, but tries to decode the information from {\bf S}. In this case, the channel capacity of the suspicious link from {\bf S} to {\bf D}, which is also  assumed to be the transmission rate by {\bf S}, is given by\footnote{The results in this paper can be readily extended to the practical scenario with non-Gaussian signaling \cite{56}, by e.g., inserting a gap $\Gamma$ in the capacity formula  \eqref{eq:RD} as $R_D=\log_2\left(1+\frac{P_S|h_{SD}|^2}{\Gamma\sigma^2} \right)$, where $\Gamma>1$ accounts for the capacity loss due to practical modulation and coding in use.}
\begin{align}
R_D=\log_2\left(1+\frac{P_S|h_{SD}|^2}{\sigma^2} \right),\label{eq:RD}
\end{align}
where $h_{SD}$ is the complex-valued channel gain from {\bf S} to {\bf D}, $P_S$ is the transmit power by {\bf S}, and $\sigma^2$ is the power of the additive white Gaussian noise (AWGN) at {\bf D}.
On the other hand, the  capacity of the single-input multiple-output (SIMO) eavesdropping channel from {\bf S} to {\bf E} is
\begin{align}
R_E=\log_2\left(1+\frac{P_S\|\mathbf h_{SE}\|^2}{\sigma^2}\right),
\end{align}
where  $\mathbf h_{SE}\in \mathbb{C}^{M\times 1}$ denotes the SIMO channel from {\bf S} to the $M$ receiving antennas of {\bf E}. If $R_E\geq R_D$ or equivalently $\|\mathbf h_{SE}\|^2\geq |h_{SD}|^2$, i.e., the legitimate monitor has a better channel than the suspicious destination, {\bf E} can reliably decode the information sent by {\bf S} with arbitrarily small error probability. As a result, the effective {\it eavesdropping rate} is given by $R_{\lk}=R_D$. On the other hand, if $R_E<R_D$, or the legitimate monitor has a weaker channel than the suspicious destination, then it is impossible for {\bf E} to decode the information from {\bf S} without any error. In this case, we define the effective   eavesdropping rate as $R_{\lk}=0$.\footnote{Note that in this case {\bf E} may still extract useful information from its received signal. However, in this paper we consider a more stringent requirement that the message from {\bf S} needs to be decoded at {\bf E} with arbitrarily small error probability to achieve the wireless surveillance goal.} Therefore, the effective eavesdropping rate can be expressed as
\begin{align} \label{eq:Rlk}
R_{\lk}=\begin{cases}
R_D, & \text{if } R_E\geq R_D \\
0, & \text{otherwise}.
\end{cases}
\end{align}

\begin{figure}
\centering
\includegraphics[scale=0.8]{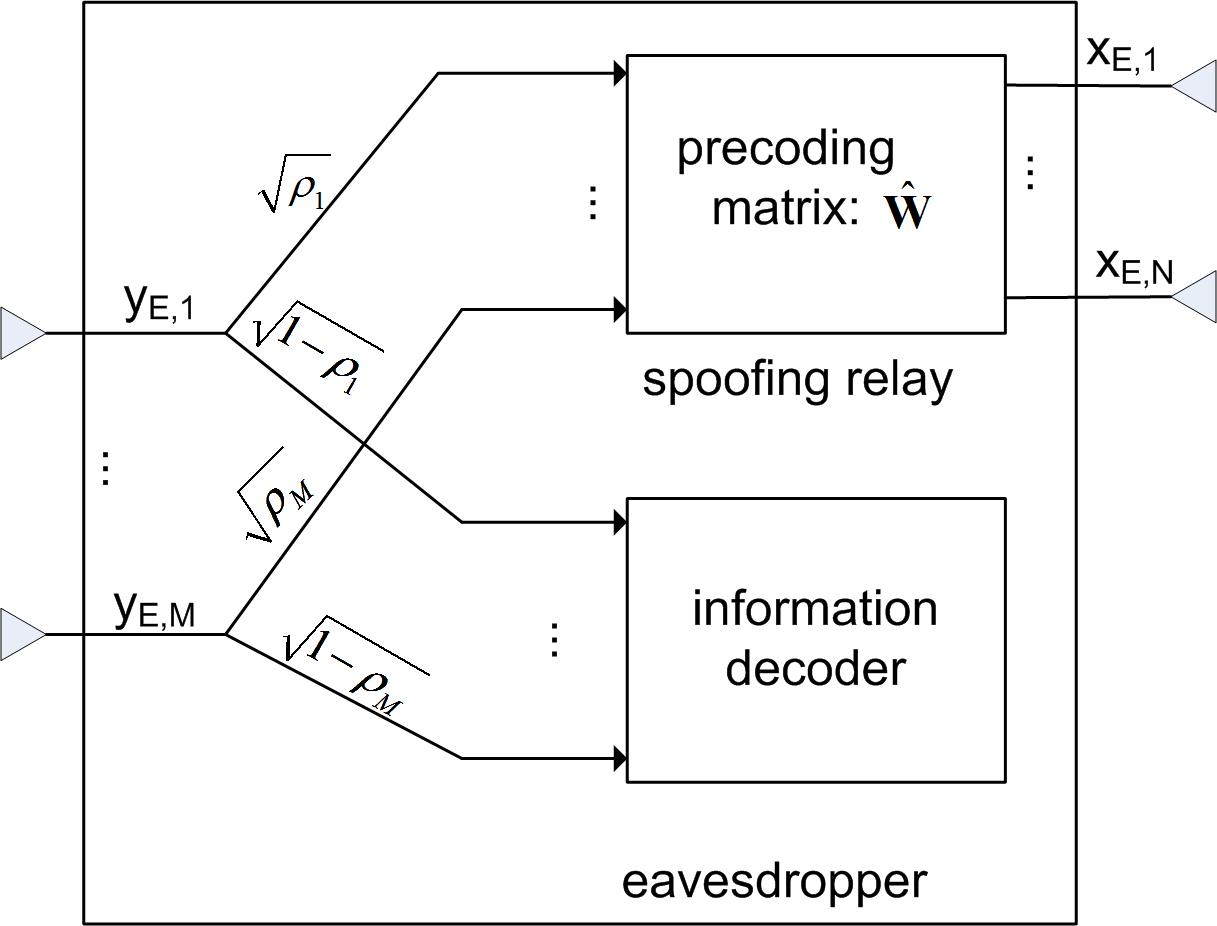}
\caption{The architecture of a multi-antenna eavesdropper with spoofing relay.}\label{F:architecture}
\end{figure}

\subsection{Proactive Eavesdropping via Spoofing Relay}
In this subsection, we propose a proactive scheme for the legitimate monitor via the spoofing relay technique to  enhance the eavesdropping rate over passive eavesdropping.
 {  To enable continuous information surveillance with concurrent proactive spoofing relaying, we assume that {\bf E} operates in a full-duplex mode with simultaneous information reception and spoofing signal transmission. As will become clear later, one key requirement for the spoofing relay technique is that the excessive time delay of the two effective signal paths (i.e., the direct link from the source to destination and that via the spoofing relay) is much smaller than the symbol duration to avoid causing inter-symbol interference (ISI). Therefore, we assume that the amplify-and-forward (AF) relaying strategy is adopted by {\bf E} since it usually has smaller processing delay than other relay processing technique such as  decode-and-forward (DF).} 
Therefore, the received signal $\mathbf y_E\in \mathbb{C}^{M\times 1}$  by the $M$ receiving antennas of {\bf E} can be expressed as
 \begin{align}
 \mathbf y_E=\mathbf h_{SE}\sqrt{P_S}d_S+ \mathbf H_{EE} \mathbf x_E + \mathbf n_E^{(A)}, \label{eq:yE}
  \end{align}
  where $d_S\sim \mathcal{CN}(0,1)$ denotes the circularly-symmetric complex Gaussian (CSCG) distributed information-bearing symbol sent by {\bf S} with transmit power $P_S$, {  $\mathbf H_{EE}\in \mathbb{C}^{M\times N}$ represents the loop channel from the $N$ transmitting antennas of {\bf E} to its own $M$ receiving antennas, $\mathbf x_E\in \mathbb{C}^{N\times 1}$ denotes the transmitted signal  by {\bf E}, and $\mathbf n_E^{(A)}\in \mathbb{C}^{M\times 1}$ represents the antenna noise received by {\bf E}. Note that the second term in the expression of $\mathbf y_E$ in  \eqref{eq:yE} is due to the full-duplex operation at {\bf E}, which in general couples the input and output signals of {\bf E}.} To avoid circuit oscillations in practice as well as to suppress the self-interference from the loop channel, the input and output of {\bf E} must be sufficiently isolated \cite{612}. 
   For ease of exposition, we assume that the ideal input-output isolation is achieved at {\bf E} by designing $\mathbf x_E$ that completely nulls the output of the loop-channel, i.e.,
  \begin{align}
  \mathbf H_{EE} \mathbf x_E = \mathbf 0.\label{eq:ZF}
  \end{align}
  Thus, the received signal by {\bf E} in \eqref{eq:yE} reduces to $\mathbf y_E=\mathbf h_{SE}\sqrt{P_S}d_S+ \mathbf n_E^{(A)}$.
   { As shown in Fig.~\ref{F:architecture}, to achieve simultaneous spoofing relaying and information eavesdropping at {\bf E} under the AF operation, the received signal $y_{E,m}$ at each antenna $m$ of {\bf E} is split into two parts, one for constructive/destructive information relaying aiming to enhance/degrade the effective channel of the suspicious link from {\bf S} to {\bf D}, and the other for information decoding so as to eavesdrop the message sent by {\bf S}.} 
   Denote by $0\leq \rho_m \leq 1$ the power splitting ratio of antenna $m$, the signal vector split for information relaying can be expressed as
  \begin{align}
  \mathbf y'_E=\diag(\sqrt{\boldsymbol \rho})\mathbf y_E= \diag(\sqrt{\boldsymbol \rho}) \left(\mathbf h_{SE}\sqrt{P_S}d_S +\mathbf n_E^{(A)}\right),
  \end{align}
  where $\boldsymbol \rho=[\rho_1, \cdots, \rho_M]^T$ represents the power splitting vector, and $\sqrt{\boldsymbol \rho}$ represents a vector obtained by taking element-wise square root. The transmitted signal $\mathbf x_E\in \mathbb{C}^{N\times 1}$ by the $N$ transmitting antennas of {\bf E} can then be expressed as
\begin{align}
\mathbf x_E&= \hat{\mathbf W}\left( \mathbf y'_E + \mathbf n_E^{(R)}\right) \label{eq:xE0} \\
&\approx \hat{\mathbf W}\left(\diag(\sqrt{\boldsymbol \rho}) \mathbf h_{SE}\sqrt{P_S}d_S + \mathbf n_E^{(R)}\right),\label{eq:xE}
\end{align}
where $\hat{\mathbf W}\in \mathbb{C}^{N\times M}$ is the relay precoding matrix at {\bf E}, and $\mathbf n_E^{(R)}\sim \mathcal{CN}(\mathbf 0,\sigma^2 \mathbf I_M)$ denotes the processing noise introduced during the relaying operation at {\bf E}, which is assumed to dominate over the antenna noise $\mathbf n_E^{(A)}$; thus, $\mathbf n_E^{(A)}$ is ignored in \eqref{eq:xE}. The transmit power by {\bf E} is thus given by
\begin{align}
\xE\left[\|\mathbf x_E\|^2 \right]=P_S\|\hat{\mathbf W} \diag(\sqrt{\boldsymbol \rho}) \mathbf h_{SE} \|^2 + \sigma^2 \|\hat{\mathbf W}\|_F^2.
\end{align}
By assuming that the processing delay due to the AF relaying at  {\bf E} is much smaller than the symbol duration, and hence is negligible, the signal received at {\bf D} can be expressed as
\begin{align}
y_D  = & h_{SD}\sqrt{P_S}d_S+\mathbf h_{ED}^H \mathbf x_E+n_D\\
  = & \underbrace{\big( h_{SD}+ \mathbf h_{ED}^H\hat{\mathbf W} \diag(\sqrt {\boldsymbol \rho})\mathbf h_{SE}\big)}_{\tilde h_{SD}}  \sqrt{P_S}d_S \notag \\
& + \underbrace{\mathbf h_{ED}^H \hat{\mathbf W} \mathbf n_E^{(R)}+n_D}_{\tilde n_D}, \label{eq:yD}
\end{align}
where $\mathbf h_{ED}^H\in \mathbb{C}^{1\times N}$ denotes the multiple-input single-output (MISO) channel from the $N$ transmitting antennas of {\bf E} to {\bf D}, and $n_D\sim \mathcal{CN}(0,\sigma^2)$ is the AWGN at {\bf D}. It is observed from \eqref{eq:yD} that by adjusting the power splitting ratio vector $\boldsymbol \rho$ and the precoding matrix $\hat{\mathbf W}$, the legitimate monitor {\bf E} is able to alter the effective channel  $\tilde{h}_{SD}$ of the suspicious link from {\bf S} to {\bf D}. The capacity of the suspicious link is thus given by $\tilde{R}_{D} = \log_2(1+\tilde{\gamma}_D)$, where $\tilde{\gamma}_D$ is the effective SNR at {\bf D}, which can be obtained from \eqref{eq:yD} as a function of $\boldsymbol \rho$ and $\hat{\mathbf W}$ as
\begin{align}
\tilde{\gamma}_D(\boldsymbol \rho, \hat{\mathbf W})=\frac{\left| h_{SD}+ \mathbf h_{ED}^H\hat{\mathbf W}\diag(\sqrt {\boldsymbol \rho})\mathbf h_{SE} \right|^2 \Ptd_S}{1+\|\mathbf h_{ED}^H \hat{\mathbf W}\|^2}, \label{eq:gammatdD}
\end{align}
where $\Ptd_S\triangleq P_S/\sigma^2$ represents the transmit SNR by {\bf S}. Note that the term $\|\mathbf h_{ED}^H \hat{\mathbf W}\|^2$ in the denominator of \eqref{eq:gammatdD} is due to the noise amplification by {\bf E}.

On the other hand, at the information decoder of {\bf E}, the split signal based on which the message from {\bf S} is decoded can be expressed as
\begin{align}
\tilde{\mathbf y}_E &=\diag(\sqrt{\boldsymbol 1-\boldsymbol \rho}) \mathbf h_{SE}\sqrt{P_S}d_S + \mathbf n_E^{(D)},
\end{align}
where $\mathbf n_E^{(D)}\sim \mathcal{CN}(\mathbf 0,\sigma^2\mathbf I_M)$ denotes the AWGN at the information decoder of {\bf E}. Thus, the  information rate achievable by {\bf E} with the optimal maximal ratio combining (MRC)  over all the $M$ signal branches  is $\tilde{R}_{E}=\log_2(1+\tilde{\gamma}_E)$, where $\tilde{\gamma}_E$ is the  SNR as a function of $\boldsymbol \rho$ given by
\begin{align}\label{eq:gammatdE}
\tilde{\gamma}_E(\boldsymbol \rho)=& \mathbf h_{SE}^H \diag(\boldsymbol 1-\boldsymbol \rho) \mathbf h_{SE}\Ptd_S\notag \\
=&\left(\|\mathbf h_{SE}\|^2-\mathbf h_{SE}^H \diag(\boldsymbol \rho) \mathbf h_{SE}\right)\Ptd_S.
\end{align}

{  To study the fundamental performance limit achievable by the legitimate monitor, we assume that perfect channel state information (CSI) of all links is available at {\bf E}. Note that in practice, the loop channel $\mathbf H_{EE}$ can be estimated beforehand at the legitimate monitor. On the other hand, the channels $\mathbf h_{SE}$ and $\mathbf h_{ED}$ could be estimated at the legitimate monitor by overhearing the pilot signals sent by {\bf S} and {\bf D}, respectively. For the suspicious link channel $h_{SD}$, it could be obtained by overhearing the channel feedback sent from {\bf D} to {\bf S}.} 

\subsection{Problem Formulation}
 The objective of the legitimate monitor {\bf E} is to jointly optimize the power splitting ratio vector $\boldsymbol \rho$ and the relay precoding matrix $\hat{\mathbf W}$ so that the eavesdropping rate is maximized. Based on the definition in \eqref{eq:Rlk}, the problem can be formulated as
\begin{align}
\mathrm{(P1)}:
\begin{cases}
  \underset{\hat{\mathbf W}, \boldsymbol \rho}{\max}  &   \ \tilde{R}_D(\boldsymbol \rho, \hat{\mathbf W})   \\
\text{ s.t. }  &  \tilde{R}_E(\boldsymbol \rho) \geq \tilde{R}_D (\boldsymbol \rho, \hat{\mathbf W}) \\
& \mathbf H_{EE} \hat{\mathbf W}=\mathbf 0,\\
& \boldsymbol 0\preceq \boldsymbol \rho \preceq \boldsymbol 1, \\
& \Ptd_S\|\hat{\mathbf W} \diag(\sqrt{\boldsymbol \rho}) \mathbf h_{SE} \|^2 + \|\hat{\mathbf W}\|_F^2 \leq \Ptd_E,
\end{cases}
\end{align}
where the zero-forcing (ZF) constraint $\mathbf H_{EE} \hat{\mathbf W}=\mathbf 0$ follows from \eqref{eq:ZF} and \eqref{eq:xE0},  and $\Ptd_E$ represents the maximum power available at {\bf E} normalized by the noise power $\sigma^2$. (P1) is a non-convex optimization problem. However, by exploiting its structure, the optimal solution  can be efficiently obtained, as shown next.

\section{Optimal Solution}\label{eq:optSol}
To obtain the optimal solution to (P1), we first consider the ZF constraint $\mathbf H_{EE} \hat{\mathbf W}=\mathbf 0$. This implies that the precoding matrix $\hat{\mathbf W}$ must lie in the null space of $\mathbf H_{EE}$. Let the (reduced) singular value decomposition (SVD) of $\mathbf H_{EE}$ be expressed as $\mathbf H_{EE}= \mathbf U_1 \boldsymbol  \Lambda_1 \mathbf V_1^H$, where $\mathbf U_1\in \mathbb{C}^{M\times r_1}$ and $\mathbf V_1\in \mathbb{C}^{N\times r_1}$ contain the $r_1$ orthonormal left and right singular vectors of $\mathbf H_{EE}$, respectively, with $r_1\leq \min\{M,N\}$ denoting the matrix rank of $\mathbf H_{EE}$, and $\boldsymbol  \Lambda_1=\diag(\lambda_1,\cdots, \lambda_{r_1})$ contains the $r_1$ positive singular values of $\mathbf H_{EE}$. Further denote by $\mathbf V_0 \in \mathbb{C}^{N\times r_0}$ with $r_0\triangleq N-r_1$ the orthogonal complement of $\mathbf V_1$, i.e., the concatenated matrix $\mathbf V\triangleq \left[ \mathbf V_1 \  \mathbf V_0\right]$ forms an orthonormal basis for the $N$-dimensional space with $\mathbf V^H \mathbf V = \mathbf I_N$. The precoding matrix $\hat{\mathbf W}$ satisfying the ZF constraint in (P1) can then be expressed as
\begin{align}
\hat{\mathbf W}= \mathbf V_0 \mathbf W, \label{eq:W}
\end{align}
where $\mathbf W\in \mathbb{C}^{r_0\times M}$ denotes the new matrix to be designed. It is not difficult to observe that in order for the ZF constraint to be feasible, we must have $r_0\geq 1$, or equivalently the rank of $\mathbf H_{EE}$ must satisfy $r_1<N$. Since $r_1\leq \min\{M,N\}$, one sufficient (but not necessary) condition is thus $N>M$, i.e., more antennas at {\bf E} should be allocated for transmission than that for reception.

By substituting \eqref{eq:W} into \eqref{eq:gammatdD}, the effective SNR at {\bf D} as a function of $\boldsymbol \rho$ and $\mathbf W$ can be expressed as
\begin{align}
\tilde{\gamma}_D(\boldsymbol \rho, \mathbf W)=\frac{\left| h_{SD}+ \hat{\mathbf h}_{ED}^H \mathbf W\diag(\sqrt {\boldsymbol \rho})\mathbf h_{SE} \right|^2 \Ptd_S}{1+\|\hat{\mathbf h}_{ED}^H \mathbf W\|^2}, \label{eq:gammatdD2}
\end{align}
where $\hat{\mathbf h}_{ED}\triangleq \mathbf V_0^H \mathbf h_{ED}$ denotes the projected channel of $\mathbf h_{ED}$ onto the null space of $\mathbf H_{EE}$. Furthermore, since the link capacity $\tilde{R}_D$ and $\tilde{R}_E$ monotonically increase with the SNR  $\gammatd_D(\boldsymbol \rho, \mathbf W)$ and $\gammatd_E(\boldsymbol \rho)$, respectively, (P1) thus reduces to
\begin{align}
\mathrm{(P2)}:
\begin{cases}
  \underset{\mathbf W, \boldsymbol \rho}{\max}  &   \ \gammatd_D(\boldsymbol \rho, \mathbf W)   \\
\text{ s.t. }  &  \gammatd_E(\boldsymbol \rho) \geq \gammatd_D(\boldsymbol \rho, \mathbf W)  \\
& \boldsymbol 0\preceq \boldsymbol \rho \preceq \boldsymbol 1, \\
& \Ptd_S\|\mathbf W \diag(\sqrt{\boldsymbol \rho}) \mathbf h_{SE} \|^2 + \|\mathbf W\|_F^2 \leq \Ptd_E,
\end{cases}
\end{align}
where the last constraint follows from the equality $\|\hat{\mathbf W} \mathbf a\|^2=\|\mathbf W \mathbf a\|^2$, $\forall \mathbf a\in \mathbb{C}^{M\times 1}$, and $ \|\hat{\mathbf W}\|_F^2=\|\mathbf W\|_F^2$ due to the fact that $\mathbf V_0^H \mathbf V_0=\mathbf I$. { Note that (P2) is equivalent to (P1) in the sense that they have the same optimal value (except the logarithmic transformation between rate and SNR), and their optimal solutions are related by the simple linear transformation equation \eqref{eq:W}. (P2) is still a non-convex optimization problem, due to the non-concave objective function as well as the non-convexity of the first constraint. However, by exploiting the special structure of the SNR expressions for $\gammatd_E(\boldsymbol \rho)$ and $\gammatd_D(\boldsymbol \rho, \mathbf W)$, the optimal solution to (P2) can be efficiently obtained. Specifically, by firstly optimizing the precoding matrix $\mathbf W$ with fixed power splitting vector $\boldsymbol \rho$, the maximum and minimum achievable SNRs (or equivalently the achievable SNR range) at {\bf D} with fixed $\boldsymbol \rho$ can be obtained. The problem (P2) then reduces to finding the optimal power splitting vector $\boldsymbol \rho$, as formulated in (P3) in Section~\ref{sec:optimalSol}. The details are given next.}

\subsection{Maximum Achievable SNR at {\bf D} with Fixed $\boldsymbol \rho$}
To obtain the optimal solution to (P2), it is noted that the SNR $\gammatd_E(\boldsymbol \rho)$ at {\bf E} only depends on the power splitting ratio vector $\boldsymbol \rho$, rather than the precoding matrix $\mathbf W$. Thus, for any fixed $\boldsymbol \rho$, we first obtain the maximum achievable SNR at {\bf D}, denoted as $\gammatd_D^{\max}(\boldsymbol \rho)$, by optimizing $\mathbf W$  as
\begin{align}
\gammatd^{\max}_D(\boldsymbol \rho)\triangleq
\begin{cases}\label{eq:rDMax}
 \underset{\mathbf W}{\max } &\  \gammatd_D(\boldsymbol \rho, \mathbf W) \\
\text{s.t.} & \Ptd_S\|\mathbf W \hath_{SE} \|^2 + \|\mathbf W\|_F^2 \leq \Ptd_E,
\end{cases}
\end{align}
where we have defined $\hath_{SE}\triangleq \diag(\sqrt{\boldsymbol \rho}) \mathbf h_{SE}$.

\begin{theorem}\label{theo:theo1}
The optimal solution to problem \eqref{eq:rDMax} is
\begin{align}
\mathbf W^\star_1 = \sqrt{\mu^\star_1} e^{j\angle h_{SD}} \tdh_{ED} \tdh_{SE}^H, \label{eq:Wopt1}
\end{align}
where $\tdh_{ED}\triangleq \hath_{ED}/\|\hath_{ED}\|$, $\tdh_{SE}\triangleq \hath_{SE}/\|\hath_{SE}\|$, and
$\mu^\star_1=\min\left\{\frac{\|\hat{\mathbf h}_{SE}\|^2}{|h_{SD}|^2\|\hat{\mathbf h}_{ED}\|^2},  \frac{\Ptd_E}{\Ptd_S \|\hat{\mathbf h}_{SE}\|^2 + 1}\right\}$.
Furthermore, the corresponding optimal value is given by \eqref{eq:gammaDMax} shown at the top of the next page.
\begin{figure*}
\begin{align}
\gammatd^{\max}_D(\boldsymbol \rho)=
\begin{cases}
\left(1+\frac{\|\hath_{SE}\|^2}{|h_{SD}|^2} \right)\Ptd_S|h_{SD}|^2, & \text{ if } \frac{\|\hat{\mathbf h}_{SE}\|^2}{|h_{SD}|^2\|\hat{\mathbf h}_{ED}\|^2} \leq  \frac{\Ptd_E}{\Ptd_S \|\hat{\mathbf h}_{SE}\|^2 + 1} \\
\frac{\left(\sqrt{\Ptd_S\|\hath_{SE}\|^2+1}+\sqrt{\Ptd_E \|\hath_{SE}\|^2\|\hath_{ED}\|^2/|h_{SD}|^2}\right)^2 \Ptd_S|h_{SD}|^2}{\Ptd_S\|\hath_{SE}\|^2+\Ptd_E\|\hath_{ED}\|^2+1}, & \text{ otherwise}. \label{eq:gammaDMax}
\end{cases}
\end{align}
\hrule
\end{figure*}
\end{theorem}
\begin{IEEEproof}
Please refer to Appendix~\ref{A:theo1}.
\end{IEEEproof}

Theorem~\ref{theo:theo1} shows that in order to maximize the  SNR at {\bf D}, the precoding matrix $\mathbf W$ at {\bf E} should be chosen such that the two signal paths from {\bf S} to {\bf D}, namely the direct link and that via the monitor relaying,  add constructively, i.e., $\angle h_{SD} = \angle \hat{\mathbf h}_{ED}^H \mathbf W\hath_{SE}$. We term such a strategy of the spoofing relay as {\it constructive information forwarding}, since it helps enhance the effective channel of the suspicious link from {\bf S} to {\bf D}. Furthermore, \eqref{eq:Wopt1} shows that the optimal relay precoding matrix $\mathbf W^\star_1$ is given by a rank-1 matrix, with a simple MRC-based linear combining (by the term $\tdh_{SE}^H$) of the output of the power splitters cascaded by a  maximal-ratio transmission (MRT) beamforming (by the term $\tdh_{ED}$) \cite{56}.

It is also noted from \eqref{eq:gammaDMax} that for any fixed power splitting ratios $\boldsymbol \rho$, the maximum achievable SNR $\gammatd^{\max}_D(\boldsymbol \rho)$ at {\bf D} depends on $\boldsymbol \rho$ only via the term $\|\hath_{SE}\|^2=\mathbf h_{SE}^H \diag(\boldsymbol \rho)\mathbf h_{SE}$. In particular, if $\boldsymbol \rho=\boldsymbol 0$, i.e., no information forwarding is applied at {\bf E}, we have $\gammatd^{\max}_D(\boldsymbol 0)=\Ptd_S|h_{SD}|^2$, which corresponds to the special case of passive eavesdropping previously discussed in Section~\ref{sec:passive}. On the other hand, if $\boldsymbol \rho=\boldsymbol 1$ and $P_E$ is sufficiently large, we have $\gammatd^{\max}_D(\boldsymbol 1)=\Ptd_S(|h_{SD}|^2+\|\mathbf h_{SE}\|^2)$. This results in the maximum  SNR achievable at {\bf D}  which is equivalent to that achieved by  MRC-based signal detection jointly performed at {\bf E} and {\bf D}. However, in this case, there is no signal  split for information decoding at the legitimate monitor and thus the effective eavesdropping rate will be zero.

\subsection{Minimum Achievable SNR at {\bf D} with Fixed $\boldsymbol \rho$}
Next, for fixed power splitting ratios $\boldsymbol \rho$, we study the minimum achievable SNR at {\bf D}, denoted as $\gammatd_D^{\min}(\boldsymbol \rho)$, which can be obtained by solving the following optimization problem,
\begin{align}
\gammatd^{\min}_D(\boldsymbol \rho)\triangleq
\begin{cases}\label{eq:rDMin}
 \underset{\mathbf W}{\min } &\  \gammatd_D(\boldsymbol \rho, \mathbf W) \\
\text{s.t.} & \Ptd_S\|\mathbf W \hath_{SE} \|^2 + \|\mathbf W\|_F^2 \leq \Ptd_E.
\end{cases}
\end{align}

\begin{theorem}\label{theo:theo2}
The optimal solution to problem \eqref{eq:rDMin} is
\begin{align}
\mathbf W_2^\star=-e^{j\angle h_{SD}}\tdh_{ED}\left(\sqrt{z_1^\star}\tdh_{SE} + \sqrt{z_2^\star}\tdh_{SE}^\perp\right)^H, \label{eq:W2opt}
\end{align}
where $\tdh_{SE}^\perp$ is any unit-norm vector that is orthogonal to $\tdh_{SE}$, i.e., $\tdh_{SE}^H\tdh_{SE}^{\perp}=0$, and $z_1^\star$ and $z_2^\star$ are the optimal solution to the following problem,
\begin{align}
\gammatd^{\min}_D(\boldsymbol \rho)\triangleq
\begin{cases}\label{eq:rDMin3}
 \underset{z_1, z_2}{\min } &\  \gammatd''_D(z_1, z_2)\triangleq \frac{\left( |h_{SD}|- \sqrt{z_1} \|\hath_{ED}\|\|\hath_{SE}\| \right)^2 \Ptd_S}{1+\|\hat{\mathbf h}_{ED}\|^2(z_1+z_2)} \\
\text{s.t.} & \left(1+\Ptd_S\|\hath_{SE}\|^2\right)z_1 +  z_2 \leq \Ptd_E, \\
& z_1, z_2 \geq 0.
\end{cases}
\end{align}
\end{theorem}
\begin{IEEEproof}
Please refer to Appendix~\ref{A:theo2}.
\end{IEEEproof}

Theorem~\ref{theo:theo2} shows that in order to minimize the SNR at {\bf D}, the precoding matrix $\mathbf W$ by {\bf E} should be chosen such that the two effective signal paths from {\bf S} to {\bf D} add destructively, i.e., $\angle h_{SD} = \pi+ \angle \hat{\mathbf h}_{ED}^H \mathbf W\hath_{SE}$. Such a strategy at {\bf E} is termed as {\it destructive information forwarding}, which degrades the effective channel of the suspicious link from {\bf S} to {\bf D}. Furthermore, similar to $\mathbf W_1^\star$ for  maximizing SNR at {\bf D},  $\mathbf W_2^\star$ in \eqref{eq:W2opt} for SNR minimization at {\bf D} is also a rank-one matrix with a similar structure. However, although the MRT-based transmit beamforming still applies, the preceding combining vector for the power-splitting output  is in general given by a linear combination of the MRC combining $\tdh_{SE}$ and its orthogonal vector $\tdh_{SE}^\perp$. Note that although the power allocated along the direction $\tdh_{SE}^\perp$ does not forward any destructive source signal to {\bf D}, it also contributes to the SNR degradation at {\bf D}  via {\it jamming}, i.e., amplifying the noise at {\bf E} to {\bf D}. Thus, the optimal relaying strategy by {\bf E} for SNR minimization at {\bf D} in general involves both destructive information forwarding and jamming.

\begin{theorem}\label{theo:theo3}
The optimal solution and the corresponding optimal value of problem \eqref{eq:rDMin3} are respectively given by \eqref{eq:z1z2Opt} and \eqref{eq:gammatdMin} shown at the top of the next page,
\begin{figure*}
\begin{align}\label{eq:z1z2Opt}
(z_1^\star, z_2^\star)=
\begin{cases}
\left(\frac{|h_{SD}|^2}{\|\hath_{ED}\|^2\|\hath_{SE}\|^2}, 0\right), & \text{ if } \tilde{P}_E\|\hath_{ED}\|^2>\tilde{P}_S|h_{SD}|^2 \text{ and } \|\hath_{SE}\|^2 \geq  \frac{|h_{SD}|^2}{\tilde{P}_E\|\hath_{ED}\|^2-\tilde{P}_S|h_{SD}|^2},\\
\left(Z_1, Z_2\right), & \text{ if } \tilde{P}_E\|\hath_{ED}\|^2\leq \tilde{P}_S|h_{SD}|^2 \text{ and } \|\hath_{SE}\|^2 C_1> \big(1+\tilde{P}_E\|\hath_{ED}\|^2\big)^2 \\
\left(\frac{\tilde{P}_E}{1+\tilde{P}_S\|\hath_{SE}\|^2}, 0\right), & \text{ otherwise},
\end{cases}
\end{align}
\begin{align}\label{eq:gammatdMin}
\gammatd^{\min}_D(\boldsymbol \rho)=
\begin{cases}
0, & \hspace{-20ex} \text{ if } \tilde{P}_E\|\hath_{ED}\|^2>\tilde{P}_S|h_{SD}|^2 \text{ and } \|\hath_{SE}\|^2 \geq  \frac{|h_{SD}|^2}{\tilde{P}_E\|\hath_{ED}\|^2-\tilde{P}_S|h_{SD}|^2},\\
\frac{\tilde{P}_S|h_{SD}|^2}{1+\tilde{P}_E\|\hath_{ED}\|^2}-1, &  \hspace{-20ex} \text{ if } \tilde{P}_E\|\hath_{ED}\|^2\leq \tilde{P}_S|h_{SD}|^2 \text{ and } \|\hath_{SE}\|^2 C_1> \big(1+\tilde{P}_E\|\hath_{ED}\|^2\big)^2 \\
\frac{\left(\sqrt{1+\tilde{P}_S\|\hath_{SE}\|^2} -\frac{\|\hath_{ED}\|\|\hath_{SE}\|}{|h_{SD}|}\sqrt{\tilde{P}_E}\right)^2\tilde{P}_S|h_{SD}|^2}{1+\tilde{P}_S\|\hath_{SE}\|^2
+\tilde{P}_E\|\hath_{ED}\|^2}, & \text{otherwise}.
\end{cases}
\end{align}
\hrule
\end{figure*}
where
\begin{align}
& Z_1\triangleq \frac{\big(1+\tilde{P}_E\|\hath_{ED}\|^2 \big)^2}{\tilde{P}_S^2 |h_{SD}|^2 \|\hath_{ED}\|^2\|\hath_{SE}\|^2},\notag \\
&Z_2\triangleq \tilde{P}_E-(1+\tilde{P}_S\|\hath_{SE}\|^2 )Z_1,\\
&C_1\triangleq \tilde{P}_S\left(\tilde{P}_S\tilde{P}_E|h_{SD}|^2\|\hath_{ED}\|^2-\big(1+\tilde{P}_E\|\hath_{ED}\|^2\big)^2\right).
\end{align}
\end{theorem}
\begin{IEEEproof}
Please refer to Appendix~\ref{A:theo3}.
\end{IEEEproof}

The results in \eqref{eq:z1z2Opt} and \eqref{eq:gammatdMin} show that if both $\tilde{P}_E$ and the split power $\|\hath_{SE}\|^2$ for information relaying are sufficiently large (corresponding to the first case in \eqref{eq:z1z2Opt} and \eqref{eq:gammatdMin}), the destructive relaying signal from {\bf E} is able to completely cancel the signal of the direct path from {\bf S} at {\bf D}, and thus makes the SNR at {\bf D} equal to zero. In this case, no dedicated jamming is needed ($z_2^\star=0$) for SNR minimization at {\bf D}. On the other hand, if destructive information relaying is unable to drive the SNR at {\bf D} to zero with the given transmit power of {\bf E}, both destructive relaying and jamming are in general needed for the SNR degradation at {\bf D}. 
Furthermore, similar to that in \eqref{eq:gammaDMax}, the minimum achievable SNR $\gammatd^{\min}_D(\boldsymbol \rho)$ in \eqref{eq:gammatdMin} depends on $\boldsymbol \rho$ only via the term $\|\hath_{SE}\|^2$. In particular, if $\boldsymbol \rho=\mathbf 0$, i.e., no information forwarding is applied at {\bf E}, we have $\gammatd^{\min}_D(\mathbf 0)=\tilde{P}_S|h_{SD}|^2/(1+\tilde{P}_E\|\hath_{ED}\|^2)$, which corresponds to jamming (merely noise amplification) with full power of {\bf E}.

\subsection{Optimal Solution to (P2)}\label{sec:optimalSol}
Now we obtain the optimal solution to problem (P2) based on the above results. Since $\gammatd_D(\boldsymbol \rho, \mathbf W)$ given in \eqref{eq:gammatdD2} is a continuous function of $\mathbf W$, for any fixed power splitting ratio vector $\boldsymbol 0 \preceq \boldsymbol \rho \preceq \boldsymbol 1$, the set of achievable SNRs at {\bf D} is given by the interval $\left[\gammatd_D^{\min}(\boldsymbol \rho), \gammatd_D^{\max}(\boldsymbol \rho) \right]$, with $\gammatd_D^{\max}(\boldsymbol \rho)$ and $\gammatd_D^{\min}(\boldsymbol \rho)$ denoting the maximum and minimum achievable SNRs given in closed-forms by \eqref{eq:gammaDMax} and \eqref{eq:gammatdMin}, respectively. As a result, (P2) reduces to finding the optimal power splitting ratios $\boldsymbol \rho$ via solving the following optimization problem,
\begin{align}
\mathrm{(P3)}: \begin{cases}
 \underset{\boldsymbol \rho, \gammatd_D}{\max} & \ \gammatd_D \\
\text{s.t.} & \gammatd_D^{\min}(\boldsymbol \rho)\leq \gammatd_D \leq \gammatd_D^{\max}(\boldsymbol \rho) \\
& \gammatd_D \leq \gammatd_E(\boldsymbol \rho) \\
& \boldsymbol 0 \preceq \boldsymbol \rho \preceq \boldsymbol 1.
\end{cases}
\end{align}

\begin{theorem}\label{theo:thoe4}
Without loss of optimality  to (P3), the power splitting ratio vector $\boldsymbol \rho$ can be expressed as
$\boldsymbol \rho=\rho \boldsymbol 1$ for $0\leq \rho\leq 1$.
\end{theorem}
\begin{IEEEproof}
Please refer to Appendix~\ref{A:thoe4}.
\end{IEEEproof}

Theorem~\ref{theo:thoe4} shows that {\it uniform power splitting} (UPS), i.e., all the $M$ receiving antennas at {\bf E} employ the same power splitting ratio, is optimal to $\mathrm{(P3)}$. By substituting with $\boldsymbol \rho=\rho \boldsymbol 1$ and after some simple manipulations, $\gammatd_E(\boldsymbol \rho)$ in   \eqref{eq:gammatdE} and  $\gammatd_{D}^{\max}(\boldsymbol \rho)$ in \eqref{eq:gammaDMax} respectively reduce to the uni-variate functions
\begin{align}
&\gammatd_E(\rho)=(1-\rho)\|\mathbf h_{SE}\|^2\tilde{P}_S, \label{eq:gammaEUPS} \\
&\gammatd_{D}^{\max}(\rho)=
\begin{cases}
  \left(1+\frac{\rho \|\mathbf h_{SE}\|^2}{|h_{SD}|^2} \right)\Ptd_S|h_{SD}|^2,  &  \hspace{-12ex} 0\leq \rho \leq \rho_1 \\
 \frac{\left(\sqrt{1+\rho \|\mathbf h_{SE}\|^2 \Ptd_S} +\frac{\|\mathbf h_{SE}\|\|\hath_{ED}\|}{|h_{SD}|}\sqrt{\rho \Ptd_E}\right)^2 \Ptd_S\hSDsq}{1+\rho \|\mathbf h_{SE}\|^2\Ptd_S+\|\hath_{ED}\|^2\Ptd_E}, & \\
  & \hspace{-12ex} \rho_1 < \rho \leq 1  ,
\end{cases}\label{eq:gammatdMaxUPS}
\end{align}
where $\rho_1\triangleq \min \left\{1, \frac{-1+\sqrt{1+4\Ptd_S\Ptd_E|h_{SD}|^2 \|\hath_{ED}\|^2}}{2\|\mathbf h_{SE}\|^2\Ptd_S}\right\}$.
Furthermore, $\gammatd^{\min}_D(\boldsymbol \rho)$ in \eqref{eq:gammatdMin} reduces to the uni-variate function \eqref{eq:gammatdMinUPS} shown at the top of the next page,
\begin{figure*}
\begin{align}\label{eq:gammatdMinUPS}
\gammatd^{\min}_D(\rho)=
\begin{cases}
0, & \hspace{-20ex}   \rho_2< \rho \leq 1 \text{ for } \tilde{P}_E\|\hath_{ED}\|^2>\tilde{P}_S|h_{SD}|^2  \\
\frac{\tilde{P}_S|h_{SD}|^2}{1+\tilde{P}_E\|\hath_{ED}\|^2}-1, &  \hspace{-20ex}  \rho_3< \rho \leq 1 \text{ for } \tilde{P}_E\|\hath_{ED}\|^2\leq \tilde{P}_S|h_{SD}|^2  \\
\frac{\left(\sqrt{1+\rho \|\mathbf h_{SE}\|^2 \Ptd_S} -\frac{\|\mathbf h_{SE}\|\|\hath_{ED}\|}{|h_{SD}|}\sqrt{\rho \Ptd_E}\right)^2 \Ptd_S\hSDsq}{1+\rho \|\mathbf h_{SE}\|^2\Ptd_S+\|\hath_{ED}\|^2\Ptd_E},  & \text{otherwise},
\end{cases}
\end{align}
\hrule
\end{figure*}
where $\rho_2\triangleq \min\left\{1, \frac{|h_{SD}|^2}{\|\mathbf h_{SE}\|^2 (\tilde{P}_E \|\hath_{ED}\|^2-\tilde{P}_S|h_{SD}|^2)} \right\}$, and $\rho_3=C_2$ if $0\leq C_2 \leq 1$, and $\rho_3=1$ otherwise, with $C_2\triangleq \frac{(1+\tilde{P}_E\|\hath_{ED}\|^2)^2}{\|\mathbf h_{SE}\|^2 \tilde{P}_S\big(\tilde{P}_S\tilde{P}_E|h_{SD}|^2\|\hath_{ED}\|^2-(1+\tilde{P}_E\|\hath_{ED}\|^2)^2\big)}$.

\begin{figure*}%
\centering
\subfigure[Case 1: $\hSDsq < \|\mathbf h_{SE}\|^2$]{
\includegraphics[scale=0.45]{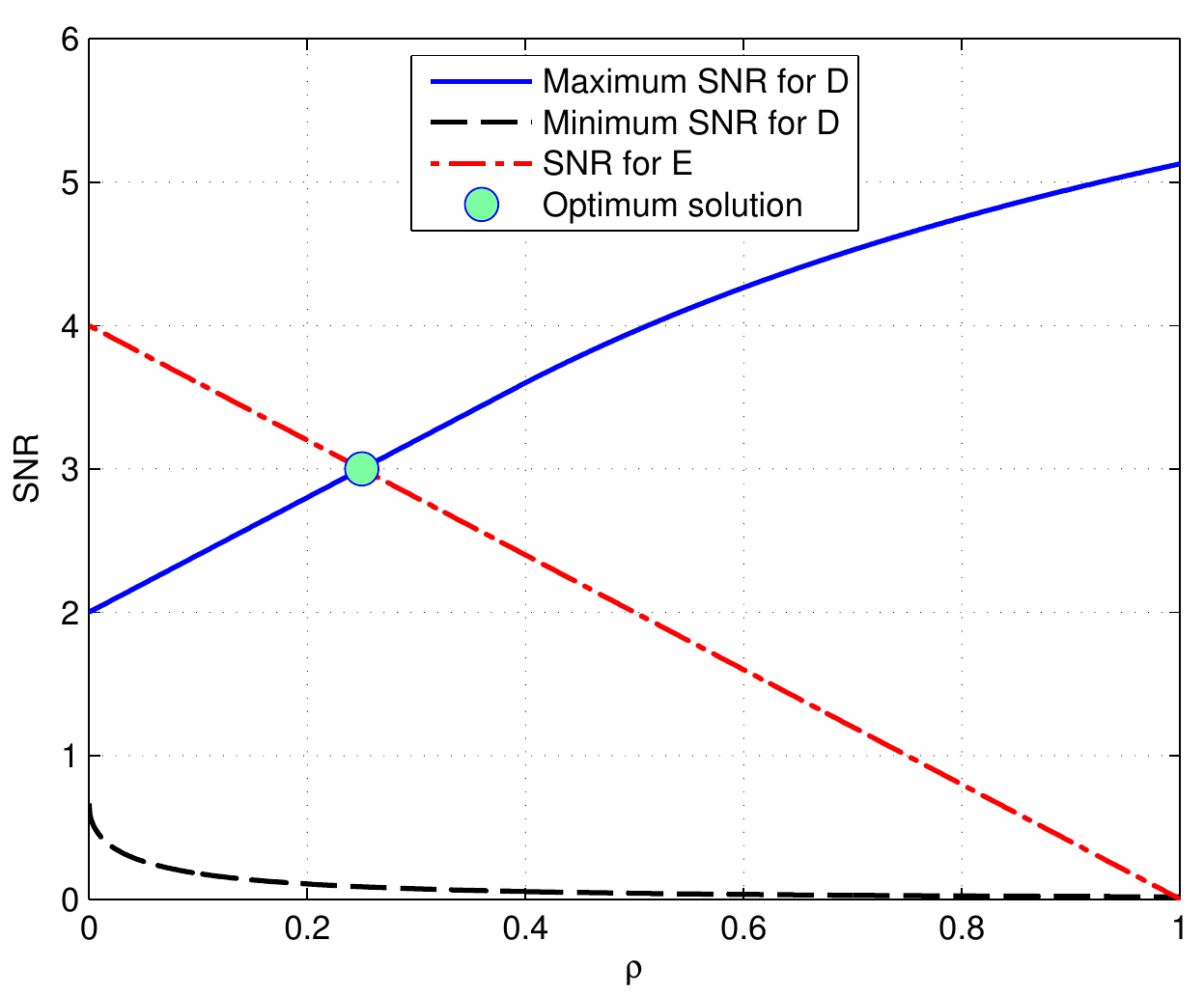}
\label{fig:subfigure1}
}
\subfigure[Case 2: $\frac{\hSDsq}{1+\|\hath_{ED}\|^2\Ptd_E}\leq \|\mathbf h_{SE}\|^2 \leq \hSDsq$]{
\includegraphics[scale=0.45]{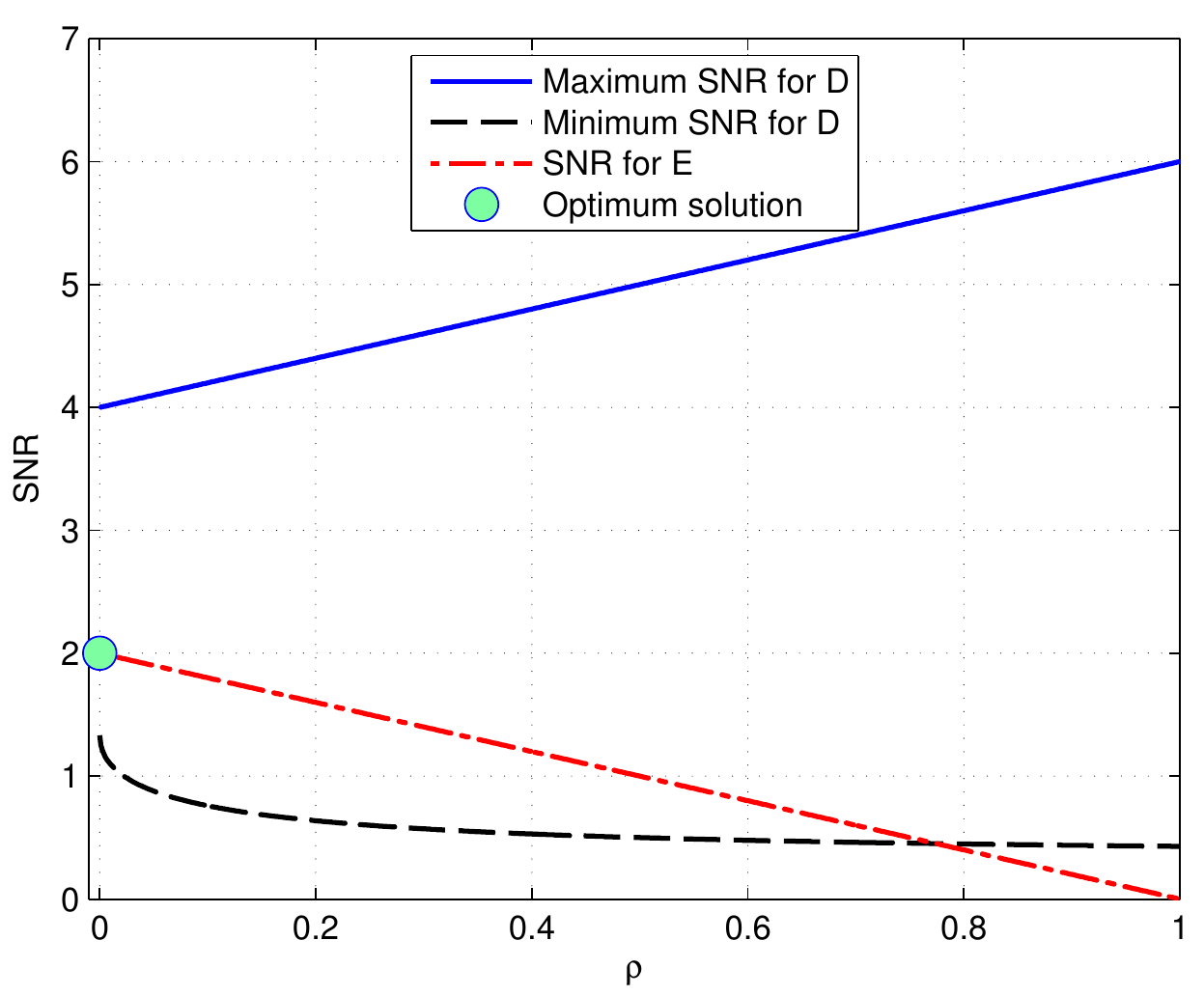}
\label{fig:subfigure2}
}
\subfigure[Case 3: $\|\mathbf h_{SE}\|^2 < \frac{\hSDsq}{1+\|\hath_{ED}\|^2\Ptd_E}$]{
\includegraphics[scale=0.45]{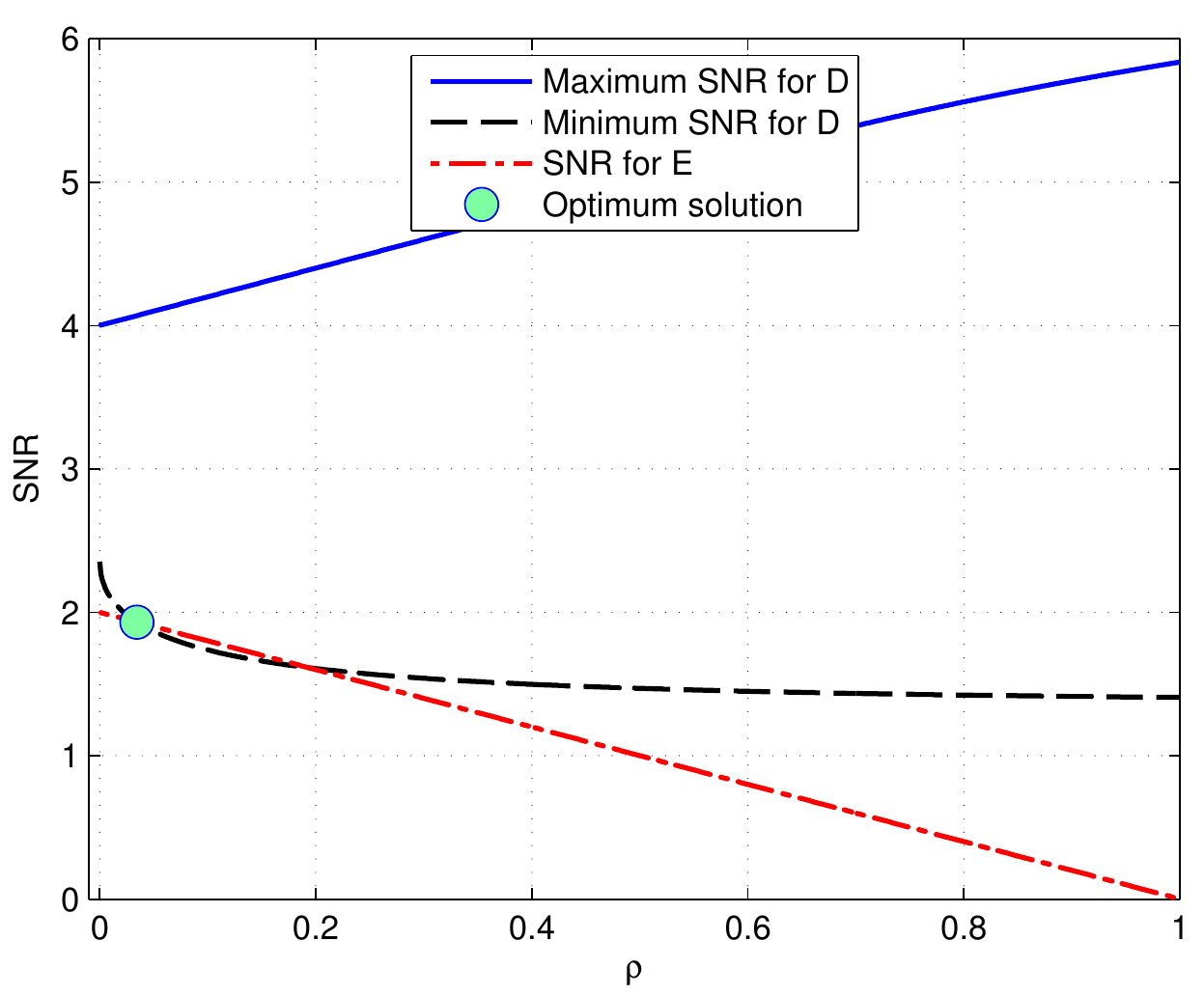}
\label{fig:subfigure3}
}
\caption{Three cases for the optimal power splitting solution.}%
\label{F:threeCases}
\end{figure*}

As a result, $\mathrm{(P3)}$ reduces to finding the optimal single power splitting ratio $\rho$ via solving
\begin{align}
\mathrm{(P3')}: \begin{cases}
 \underset{0\leq \rho \leq 1, \gammatd_D}{\max} & \ \gammatd_D \\
\text{s.t.} & \gammatd_D^{\min}(\rho)\leq \gammatd_D \leq \gammatd_D^{\max}(\rho) \\
& \gammatd_D \leq \gammatd_E(\rho).
\end{cases}
\end{align}
By noticing that $\gammatd_E(\rho)$ in \eqref{eq:gammaEUPS} is a linear decreasing function of $\rho$, $\mathrm{(P3')}$ is essentially equivalent to finding the ``best'' intersection point between the line $\gammatd_E(\rho)$ and the set
 \begin{align}
 \mathcal X \triangleq \left\{(\gamma, \rho)| 0\leq \rho \leq 1, \gammatd_D^{\min}(\rho)\leq \gamma \leq \gammatd_D^{\max}(\rho)\right\}
  \end{align}
  in the $\gamma$-$\rho$ plane, which can be optimally solved by separately considering the following three cases, as illustrated in Fig.~\ref{F:threeCases}.

 {\it Case 1: $\gammatd_D^{\max}(0)<\gammatd_E(0)$} or $\hSDsq < \|\mathbf h_{SE}\|^2$ as illustrated in Fig.~\ref{fig:subfigure1}. In this case, the legitimate monitor $\mathbf E$ has a better channel than the suspicious receiver $\mathbf D$. Intuitively, $\mathbf E$ should perform constructive information forwarding to enhance the effective channel of $\mathbf D$ so as to increase the eavesdropping rate. It follows from Fig.~\ref{fig:subfigure1} that the optimal solution to $\mathrm{(P2)}$ is given by the intersection point of the two curves $\gammatd_D^{\max}(\rho)$ and $\gammatd_E(\rho)$. As $\gammatd_D^{\max}(\rho)$ and $\gammatd_E(\rho)$  are monotonically increasing and decreasing functions over $0\leq \rho \leq 1$, respectively, and $\gammatd_D^{\max}(1)>\gammatd_E(1)=0$, the equation $\gammatd_D^{\max}(\rho)=\gammatd_E(\rho)$ has one unique solution $\rho^\star$, which can be efficiently obtained via bisection method.

Furthermore, if $\mathbf E$ has sufficiently large power $\Ptd_E$, $\rho^\star$ can be obtained in closed-form, as given in the following lemma.
\begin{lemma}\label{lemma:lm2}
If $\alpha \triangleq \frac{\|\mathbf h_{SE}\|^2}{\hSDsq}>1$ and $\Ptd_E\geq \frac{\alpha\left( 1+ \|\mathbf h_{SE}\|^2\Ptd_S\right)}{\|\hath_{ED}\|^2}$, the optimal solution to $\mathrm{(P3')}$ is $\rho^\star = \frac{1}{2}\left(1-\frac{1}{\alpha} \right)$, and the SNRs of both {\bf D} and {\bf E} are $\gammatd_D^\star=\gammatd_E^\star=\frac{\|\mathbf h_{SE}\|^2+\hSDsq}{2}\Ptd_S$.
\end{lemma}
\begin{IEEEproof}
Lemma~\ref{lemma:lm2} directly follows by noticing that with sufficiently large $\Ptd_E$ as specified in the lemma, $\gammatd_D^{\max}(\rho)$ in \eqref{eq:gammatdMaxUPS} reduces to the linear function of $\rho$ since $\rho_1=1$.
\end{IEEEproof}
 Lemma~\ref{lemma:lm2} shows that by employing constructive relaying with the optimal power splitting ratio $\rho^\star$, {\bf E} is able to increase the eavesdropping rate as compared to passive eavesdropping since $\gammatd_D^\star >\hSDsq \Ptd_S$. Furthermore, as $\alpha$ increases,  more power should be split at $\mathbf E$ for constructive relaying to enhance the suspicious link SNR. In the extreme case when $\alpha \rightarrow \infty$, i.e., the eavesdropper's link is much stronger than the suspicious user's link, half of the power of the received signal at $\mathbf E$ should be split for information relaying, and the other half for information decoding (eavesdropping).

{\it Case 2: $\gammatd_D^{\min}(0)\leq \gammatd_E(0)\leq \gammatd_D^{\max}(0)$} or $\frac{\hSDsq}{1+\|\hath_{ED}\|^2\Ptd_E}\leq \|\mathbf h_{SE}\|^2 \leq \hSDsq$, as illustrated in Fig.~\ref{fig:subfigure2}. In this case, the eavesdropping link is worse than the suspicious link, but it becomes better if jamming with full power is applied at {\bf E} to degrade the suspicious link. It follows from Fig.~\ref{fig:subfigure2} that the optimal solution to $\mathrm{(P3')}$ is $\rho^\star=0$, i.e., no information forwarding and only  jamming should be applied at {\bf E}, where the normalized jamming power is given by $\Ptd_E^\star=\frac{1}{\|\hath_{ED}\|^2}\left(\frac{\hSDsq}{\|\mathbf h_{SE}\|^2}-1 \right)$ so as to degrade the suspicious link SNR to the same level as that at $\mathbf E$. In this case, the SNR at both {\bf D} and {\bf E} is $\gamma_D^\star=\gamma_E^\star=\|\mathbf h_{SE}\|^2 \Ptd_S$.

{\it Case 3: $\gammatd_E(0)<\gammatd_D^{\min}(0)$} or $\|\mathbf h_{SE}\|^2 < \frac{\hSDsq}{1+\|\hath_{ED}\|^2\Ptd_E}$, as illustrated in Fig.~\ref{fig:subfigure3}. In this case, the legitimate monitor's link is worse than the suspicious user's link even after jamming with full power by {\bf  E}. Therefore, destructive information forwarding and jamming should be both applied at $\mathbf E$ to further degrade the suspicious link SNR.  It follows from Fig.~\ref{fig:subfigure3} that the optimal solution $\rho^\star$ to $\mathrm{(P3')}$ is obtained by solving $\gammatd_D^{\min}(\rho)=\gammatd_E(\rho)$ in the interval $0\leq \rho \leq 1$, which can be reduced to a quartic equation and hence solved efficiently. Note that if more than one solutions exist, the one with the smallest magnitude is the optimal solution. On the other hand, if no solution exists, it implies that problem $\mathrm{(P3')}$, and hence $\mathrm{(P1)}$, is infeasible, i.e., the legitimate monitor is unable to degrade the suspicious user transmission rate to be decodable by {\bf E} with its given transmit power.

\subsection{Low-Cost Implementation with One Single Power Splitter}
The solution obtained in the preceding subsection in general leads to positive power splitting ratios at all receiving antennas of {\bf E} (except for Case 2 where   jamming only is optimal), i.e., $\rho_m>0$, $\forall m$; thus, in total $M$ power splitters need to be equipped at {\bf E}, which could be costly in practice. In this subsection, we show that for practical implementation, one single power splitter is sufficient to achieve optimal eavesdropping, regardless of the number of receiving antennas $M$.
\begin{theorem}\label{theo:theo5}
There exists an optimal solution to $\mathrm{(P3)}$ such that the power splitting ratios $\{\rho_m\}_{m=1}^M$ are given by
\begin{align}\label{eq:binary}
\rho_m=\begin{cases}
\rho, & \ m=m^\star \\
0 \text{ or } 1,& \ m\neq m^\star.
\end{cases}
\end{align}
\end{theorem}
\begin{IEEEproof}
Please refer to Appendix~\ref{A:theo5}.
\end{IEEEproof}

{  Note that Appendix~\ref{A:theo5} gives a constructive proof of Theorem~\ref{theo:theo5}, where the optimal power splitting ratio vector satisfying \eqref{eq:binary} is obtained in closed-form \eqref{eq:rhoBps} based on the optimal uniform  power splitting vector to problem (P3). Therefore, in terms of computational complexity, the (semi-)binary power splitting solution in Theorem~\ref{theo:theo5} is comparable to that of the uniform power splitting solution obtained in the preceding subsection, which are both quite efficient since they only require solving either a bisection search problem (for Case 1) or a quartic equation (for Case 3). However, in terms of practical implementation, the solution given in Theorem~\ref{theo:theo5} is more cost-effective since it requires only one power splitter to be equipped at {\bf E}, instead of $M$ as for the uniform power splitting scheme.}


\begin{figure}
\centering
\includegraphics[scale=0.25]{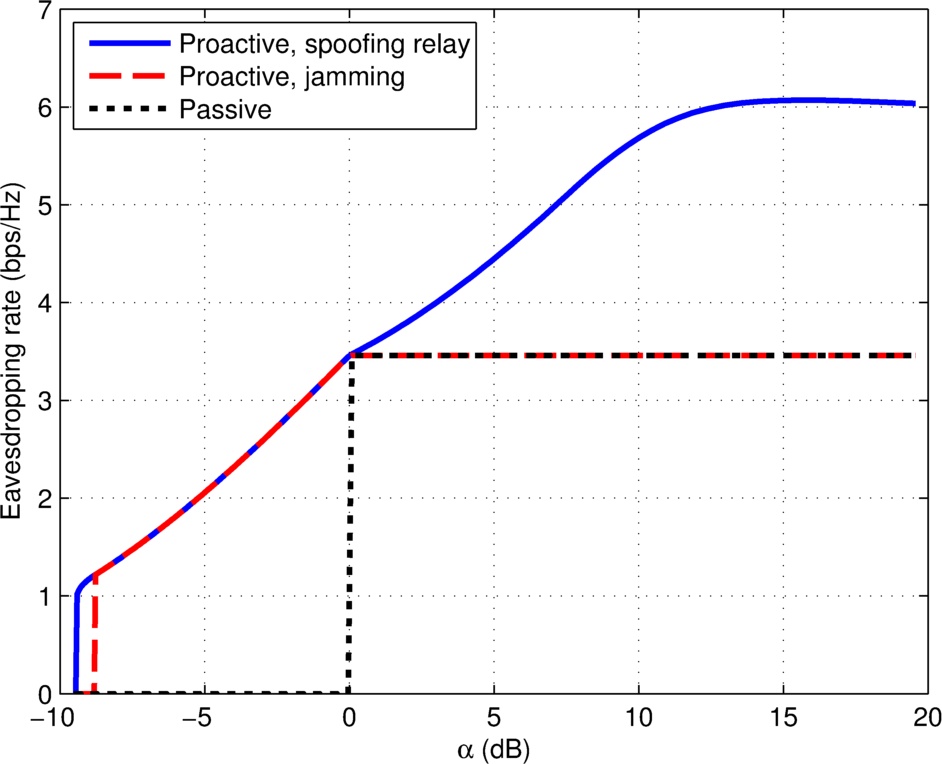}
\caption{Eavesdropping rate versus $\alpha\triangleq \|\mathbf h_{SE}\|^2/|h_{SD}|^2$, where $\gamma_0\triangleq P_S|h_{SD}|^2/\sigma^2=10$ dB and $P_E=P_S$.}\label{F:LeakageVsAlpha}
\end{figure}

\section{Numerical Results}\label{sec:numerical}
In this section, numerical results are provided to evaluate the performance of the proposed proactive eavesdropping via spoofing relay technique. {  We assume that the suspicious source $\mathbf S$ and the suspicious destination $\mathbf D$  are separated by a fixed distance $d_{SD}=1000$ meters (m). The legitimate monitor $\mathbf E$ is located on the same line connecting $\mathbf S$ to $\mathbf D$, with the distance between $\mathbf S$ and $\mathbf E$ denoted as $d_{SE}$ that varies between $100$m and $3000$m for different eavesdropper locations. Such a setup could correspond to the macro-cell users (with typical cell radius on the order of km).}  
Thus, the distance between $\mathbf E$ and $\mathbf D$ can be expressed as $d_{ED}=|d_{SD}-d_{SE}|$. We assume that uniform linear arrays with adjacent elements separated by half-wavelength are equipped at the transmitter and receiver of $\mathbf E$. Furthermore, we assume that all links are dominated by the line-of-sight (LoS) channels that follow the free-space path loss model.  Unless otherwise specified, the number of transmitting and receiving antennas at $\mathbf E$ are $N=2$ and $M=1$, respectively. The operating frequency is assumed to be $1.8$ GHz. {  We define $\alpha\triangleq \|\mathbf h_{SE}\|^2/|h_{SD}|^2$ as the channel power ratio between the eavesdropping and the suspicious links, where all the channels including $\mathbf h_{SE}$ and $h_{SD}$ are generated based on the eavesdropper's location and the LoS path loss model}. Furthermore, denote $\gamma_0\triangleq P_S |h_{SD}|^2/\sigma^2$ as the reference SNR received at $\mathbf D$ with source transmission power $P_S$ and receiver noise power $\sigma^2$. We consider two benchmark schemes, namely passive eavesdropping as discussed in Section~\ref{sec:passive}, and proactive eavesdropping with jamming only \cite{648}, \cite{644}. Note that with the definition given in \eqref{eq:Rlk}, passive eavesdropping has positive eavesdropping rate only when $\mathbf E$ has better channel than $\mathbf D$ from the suspicious source $\mathbf S$, i.e., $\alpha\geq 1$. On the other hand, jamming is helpful for enhancing the eavesdropping rate only when $\mathbf E$ has weaker channel than $\mathbf D$, i.e., $\alpha<1$.

First, we study the effect of $\alpha$ on the maximum eavesdropping rate achievable by the legitimate monitor with the three eavesdropping schemes. To this end, we assume that the legitimate monitor $\mathbf E$ moves towards $\mathbf S$ from the location with $d_{SE}=3000$m to that with $d_{SE}=100$m. Correspondingly, the channel power ratio $\alpha$ increases from around $-10$dB to $20$dB. The transmission power $P_S$ by $\mathbf S$ is fixed to a value such that the reference SNR $\gamma_0=10$ dB, and the maximum transmission power at $\mathbf E$ is set as $P_E=P_S$. The eavesdropping rates $R_\lk$ versus $\alpha$ with the three schemes are compared in Fig.~\ref{F:LeakageVsAlpha}. It is observed that when $\mathbf E$ has a better channel than $\mathbf D$, i.e., $\alpha>1$ (0 dB), both passive eavesdropping and  jamming-based eavesdropping (with zero jamming power in this case) achieve a constant $R_\lk$, which is equal to the channel capacity of the suspicious link from $\mathbf S$ to $\mathbf D$. In contrast, the proposed proactive eavesdropping scheme with spoofing relaying achieves a significantly higher eavesdropping rate, since it is able to spoof the suspicious source $\mathbf S$ to increase its transmission rate by constructively forwarding the source signal to $\mathbf D$, which is not possible in the two benchmark schemes. When $\mathbf E$ has a worse channel than $\mathbf D$, i.e., $\alpha<1$ (0 dB), the eavesdropping rate with the passive scheme drops to zero, since $\mathbf E$ cannot reliably decode the information sent from $\mathbf S$. In contrast, as long as $\alpha$ is not too small, the two proactive schemes can achieve strictly positive eavesdropping rate, since they both jam the suspicious link to spoof the source $\mathbf S$ to decrease its transmission rate to be decodable at $\mathbf E$. Note that in this regime, the proposed spoofing relay technique degenerates to jamming (Case 2 as described in Section~\ref{sec:optimalSol}), thus the two proactive schemes obtain identical rate performance. As $\alpha$  further decreases, e.g., $\alpha=-10$ dB, $\mathbf E$ is unable to decode the information sent by $\mathbf S$ even with the two proactive schemes, and thus the eavesdropping rate drops to zero. However, it is worth noting that at around $\alpha=-9$dB, the proposed spoofing relay technique still achieves strictly positive eavesdropping rate, whereas that with jamming only is zero. This corresponds to Case 3 as described in Section~\ref{sec:optimalSol}, where both jamming and destructive information forwarding can be jointly applied to further degrade the suspicious link SNR for $\mathbf E$ to decode the information sent by $\mathbf S$.

\begin{figure}
\centering
\includegraphics[scale=0.4]{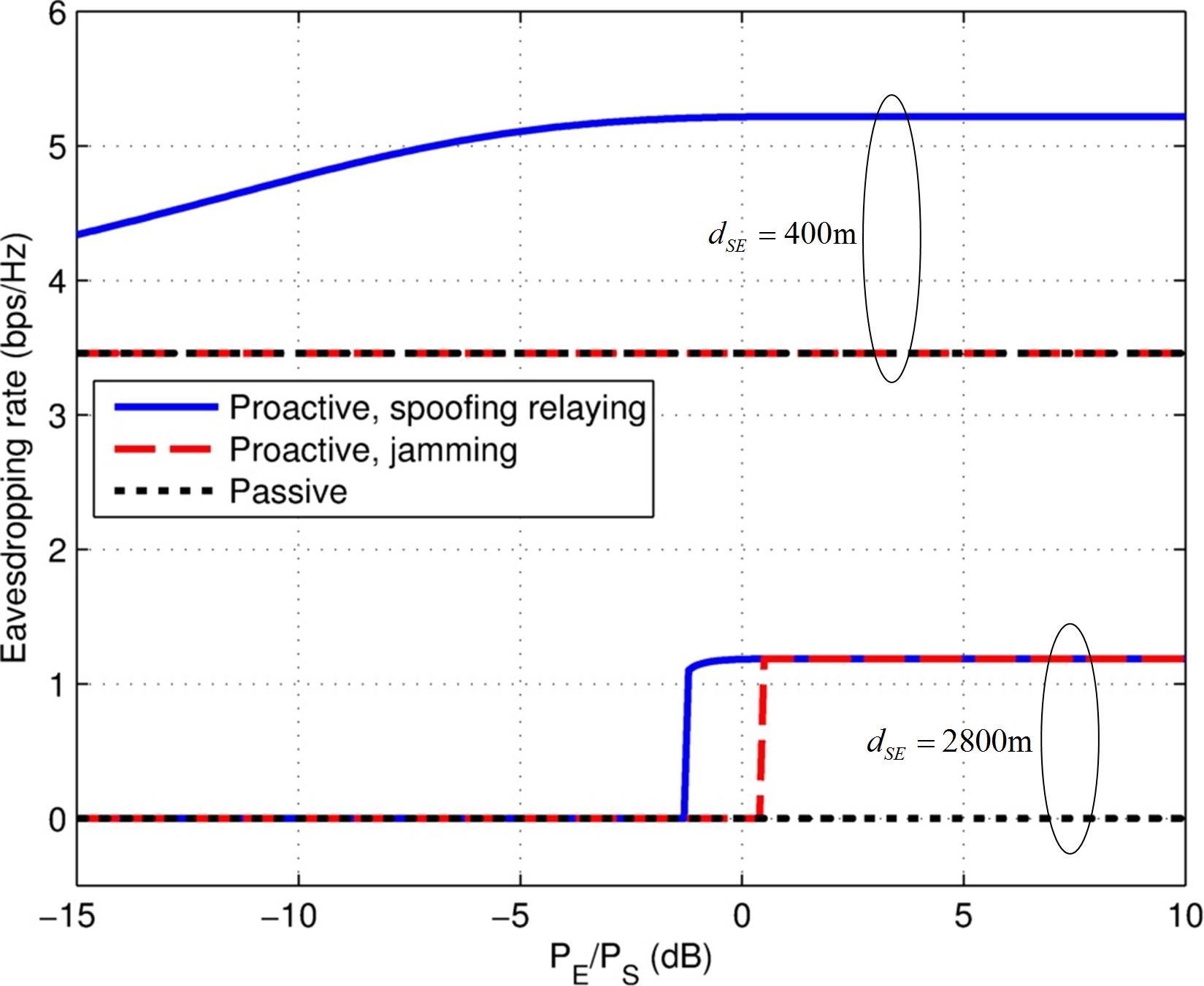}
\caption{Eavesdropping rate versus the legitimate monitor's power budget $P_E$, where $\gamma_0=10$ dB.}\label{F:LeakageVsPE}
\end{figure}

Next, we investigate the effect of the legitimate monitor's power budget $P_E$ on the maximum achievable eavesdropping rate with the three considered schemes. We compare two setups with $d_{SE}=400$m and $d_{SE}=2800$m, respectively. The transmission power $P_S$ by $\mathbf S$ is fixed such that $\gamma_0=10$ dB, whereas the power budget $P_E$ at the legitimate monitor varies such that the power ratio $P_E/P_S$ increases from $-15$ dB to $10$ dB. Fig.~\ref{F:LeakageVsPE} shows the maximum eavesdropping rate versus $P_E/P_S$ by the three eavesdropping schemes under the two setups. It is first observed that the eavesdropping rate by the passive scheme is independent of $P_E$, since no transmission is performed at $\mathbf E$. Furthermore, for the case with $d_{SE}=400$m, the eavesdropping rate by the jamming scheme is also independent of $P_E$, since no jamming is needed when the eavesdropping link is stronger than the suspicious link. In contrast, the eavesdropping rate with the proposed spoofing relay technique under $d_{SE}=400$m firstly increases with $P_E$, and then approaches to a constant value as $P_E$ gets sufficiently large, which is in accordance with Lemma~\ref{lemma:lm2}. For the case of $d_{SE}=2800$m where $\mathbf E$ has worse channel than $\mathbf D$, both the two proactive schemes have zero eavesdropping rate when $P_E$ is sufficiently small, whereas the rate increases to a positive value when $P_E$ exceeds certain thresholds. {  It is noted that if the eavesdropping link is stronger than the suspicious link (e.g., $d_{SE}=400$m), then the proposed scheme always outperforms the jamming-based scheme, regardless of the power budget $P_E$ at the relay. On the other hand, for $d_{SE}=2800$m so that the eavesdropping link is weaker than the suspicious link, an excessive relay power about $2$ dB is required by jamming than the proposed scheme in order to achieve the same eavesdropping rate of around $1.2$ bps/Hz.}

\begin{figure}
\centering
\includegraphics[scale=0.25]{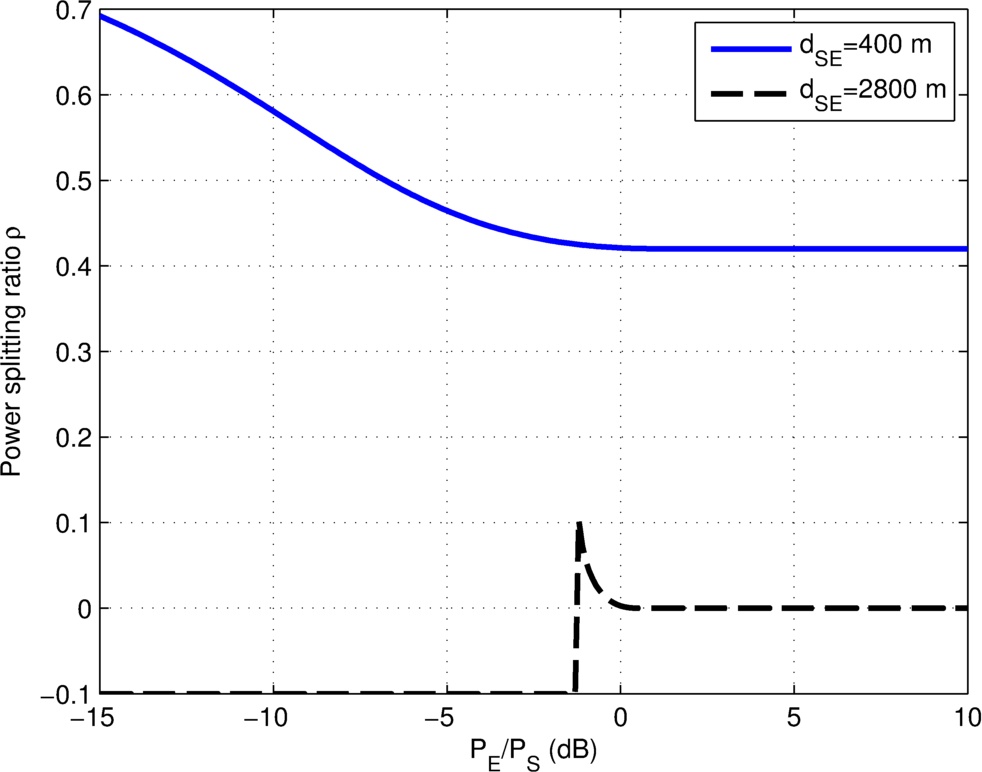}
\caption{Optimal power splitting ratio for the proposed spoofing relaying versus $P_E$, where $\gamma_0=10$ dB.}\label{F:PowerSlittingRatioVsPE}
\end{figure}

For the proposed spoofing relaying scheme, Fig.~\ref{F:PowerSlittingRatioVsPE} plots the optimal power splitting ratio $\rho^\star$ at $\mathbf E$ versus $P_E/P_S$. It is observed that for the case of $d_{SE}=400$m, $\rho^\star$ firstly decreases with $P_E$, since larger transmission power at $\mathbf E$ requires less signal power to be split for information relaying, and hence more power can be split for eavesdropping. As $P_E$ gets sufficiently large, $\rho^\star$ approaches to a constant value of $0.42$, which is in accordance with that predicted by Lemma~\ref{lemma:lm2}. On the other hand, for the case of $d_{SE}=2800$m, problem (P1) is infeasible when $P_E$ is too small, which is indicated by the infeasible $\rho=-0.1$ in Fig.~\ref{F:PowerSlittingRatioVsPE}. For $P_E/P_S$ between $-2$dB and $0$dB, a small fraction of the signal power received at $\mathbf E$ is split for destructive information forwarding. As $P_E/P_S$ increases to $0$ dB, $\rho^\star$ becomes zero since jamming alone is optimal in this case, which is consistent with the results in Fig.~\ref{F:LeakageVsPE}.

\begin{figure}
\centering
\includegraphics[scale=0.35]{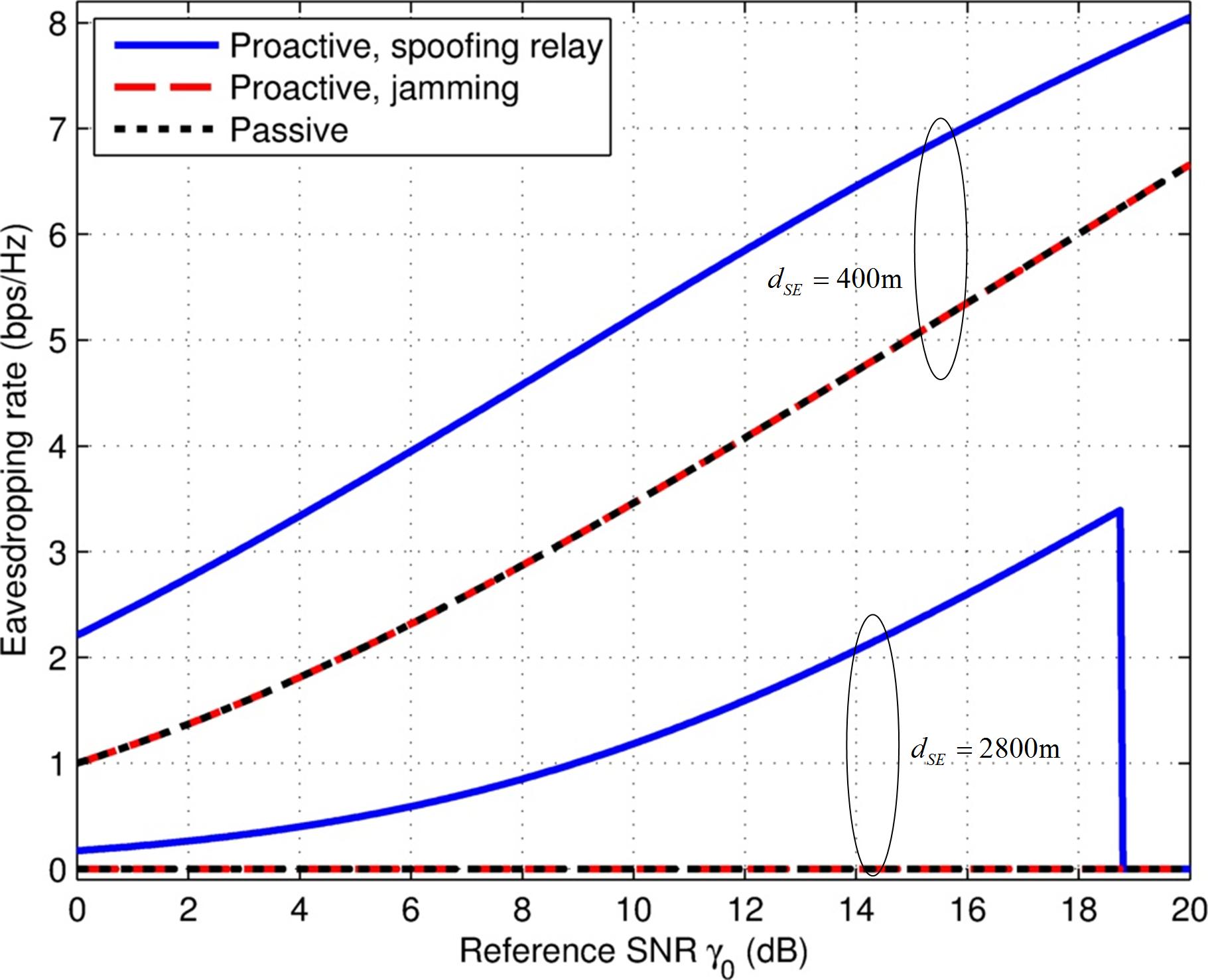}
\caption{Eavesdropping rate versus $\gamma_0\triangleq P_S|h_{SD}|^2/\sigma^2$. 
}\label{F:LeakageVsPS}
\end{figure}

Next, we investigate the effect of source transmission power $P_S$ on the eavesdropping rate. Similar to that in Fig.~\ref{F:LeakageVsPE} and Fig.~\ref{F:PowerSlittingRatioVsPE}, we consider two setups with $d_{SE}=400$m and $d_{SE}=2800$m, respectively. The  power budget $P_E$ at $\mathbf E$ is fixed such that $P_E|h_{SD}|^2/\sigma^2=10$dB, whereas the transmission power $P_S$ by the suspicious source varies such that the reference SNR $\gamma_0$ increases from $0$ dB to $20$ dB. Fig.~\ref{F:LeakageVsPS} shows the maximum achievable eavesdropping rate by the three considered schemes versus $\gamma_0$ under each of the two setups. It is first observed that for the case of $d_{SE}=400$m where $\mathbf E$ has better channel than $\mathbf D$, the eavesdropping rate with all three considered schemes increases with $P_S$. However, the proposed spoofing relaying scheme performs significantly better than the other two schemes, due to the constructive relaying performed at $\mathbf E$ that enhances the effective channel capacity from $\mathbf S$ to $\mathbf D$. For the case of $d_{SE}=2800$m, the eavesdropping rate by both passive eavesdropping and jamming only is zero due to the poor channel between $\mathbf S$ and $\mathbf E$. In contrast, the proposed spoofing relaying technique is able to achieve strictly positive eavesdropping rate for $\gamma_0$ below a certain threshold, beyond which the eavesdropping rate drops to zero since the suspicious transmission rate cannot be degraded to a value decodable by $\mathbf E$ with its given transmit power constraint $P_E$.

\begin{figure}
\centering
\includegraphics[scale=0.65]{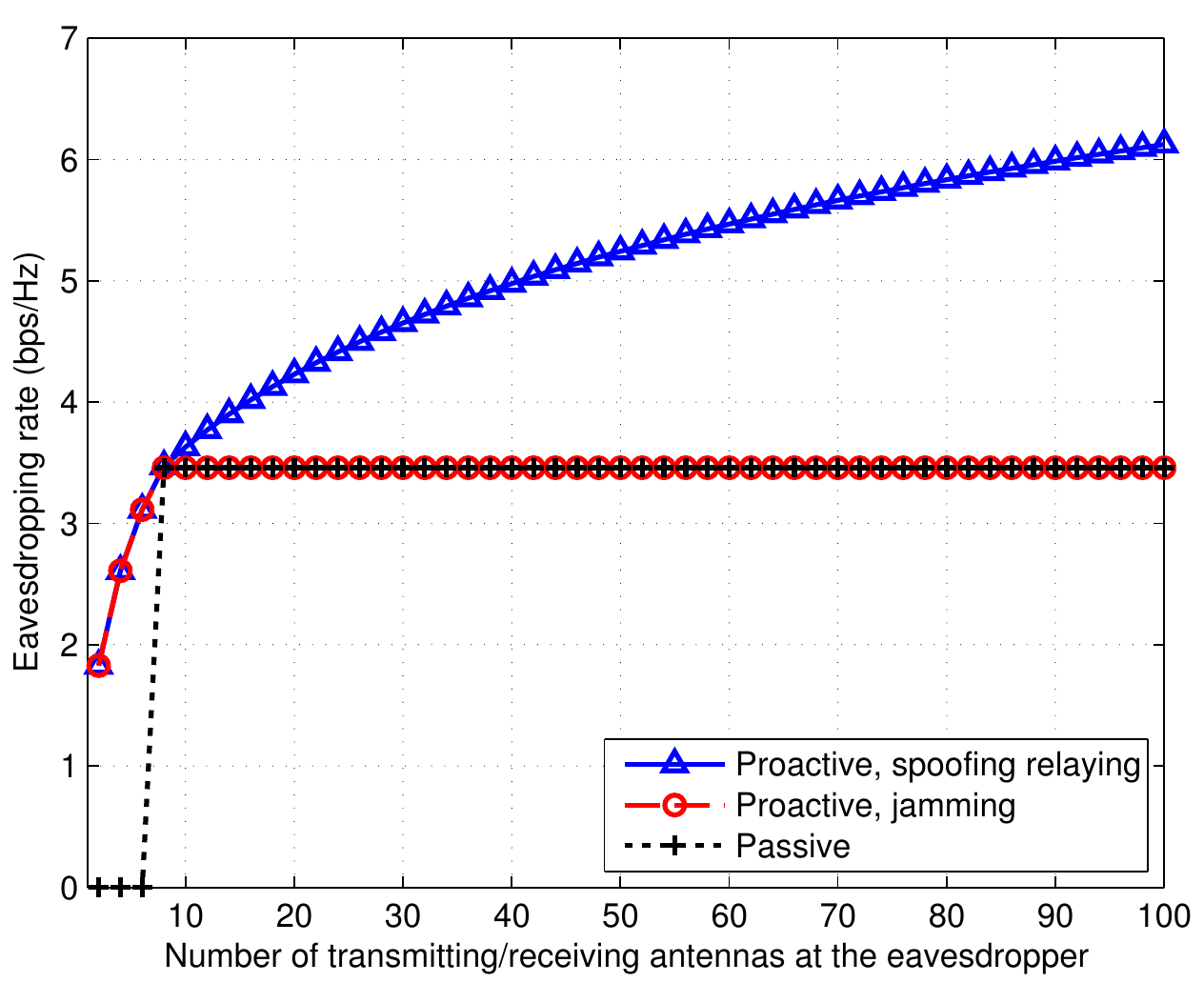}
\caption{Eavesdropping rate versus the number of transmitting/receiving antennas at the eavesdropper.
}\label{F:LeakageVsNumAnts}
\end{figure}

{  Lastly, by assuming equal number transmitting and receiving antennas at the eavesdropper, i.e., $M=N\triangleq\bar N$, Fig.~\ref{F:LeakageVsNumAnts} shows the eavesdropping rate versus $\bar N$ for $d_{SE}=2800$m, $\gamma_0=10$ dB, and $P_E=P_S$. Note that since the loop channel $\mathbf H_{EE}$ is of rank-1 under the LoS assumption, the ZF constraint \eqref{eq:ZF} is feasible as long as $N\geq 2$. Fig.~\ref{F:LeakageVsNumAnts} shows that the performance of all the three eavesdropping schemes in general improves with the increasing of $\bar N$, which is expected due to the more powerful receiving/transmitting beamforming gains as more antennas are used. However, for the two benchmark schemes, no further improvement on the eavesdropping rate is possible for $\bar N$ beyond $6$, since the eavesdropping rate is fundamentally limited by the source transmission rate, which neither passive eavesdropping nor jamming is able to improve when the eavesdropping link becomes better than the suspicious link. In contrast, thanks to the constructive relaying, the proposed spoofing relaying technique is able to achieve continuous eavesdropping rate improvement as $\bar N$ increases, though with a diminishing gain. This again shows the superior performance of the proposed eavesdropping strategy over the two benchmark schemes.
}

\section{Conclusions}\label{sec:conclusion}

{ This paper has studied the new wireless information surveillance problem in the general paradigm of wireless security. A novel proactive eavesdropping scheme with spoofing relaying technique has been proposed. With the proposed scheme, the legitimate monitor acts as a full-duplex relay for simultaneous eavesdropping and spoofing relaying  to render the suspicious source to vary transmission rate in favor of the eavesdropping performance. The receive power splitting ratios and the transmit precoding matrix  at the legitimate monitor are jointly optimized for eavesdropping rate maximization. Numerical results show that the proposed spoofing relay technique significantly enhances the information surveillance performance as compared to the two benchmark schemes with passive or jamming-based eavesdropping.}

\appendices

\section{Proof of Theorem~\ref{theo:theo1}}\label{A:theo1}
To show Theorem~\ref{theo:theo1}, we first apply the triangle inequality to \eqref{eq:gammatdD2}, which yields
\begin{align}
\tilde{\gamma}_D(\boldsymbol \rho, \mathbf W)\leq \frac{\left( \big|h_{SD}\big|+ \big| \hat{\mathbf h}_{ED}^H \mathbf W\hath_{SE}\big| \right)^2 \Ptd_S}{1+\|\hat{\mathbf h}_{ED}^H \mathbf W\|^2}, \ \forall \mathbf W,
\end{align}
where  equality holds if and only if $\mathbf W$ is chosen such that the two effective signal paths from {\bf S} to {\bf D} add constructively, i.e., $\angle h_{SD} = \angle \hat{\mathbf h}_{ED}^H \mathbf W\hath_{SE}$. Furthermore, since an arbitrary phase shifting of $\mathbf W$ does not alter the feasibility of the power constraint in \eqref{eq:rDMax}, problem \eqref{eq:rDMax} is thus equivalent to
\begin{align}
\gammatd^{\max}_D(\boldsymbol \rho)\triangleq
\begin{cases}\label{eq:rDMax2}
 \underset{\mathbf W}{\max } &\  \gammatd'_D(\mathbf W)\triangleq \frac{\left( |h_{SD}|+ \big |\hat{\mathbf h}_{ED}^H \mathbf W \hat{\mathbf h}_{SE}\big| \right)^2 \Ptd_S}{1+\|\hat{\mathbf h}_{ED}^H \mathbf W\|^2} \\
\text{s.t.} & \Ptd_S\|\mathbf W \hat{\mathbf h}_{SE} \|^2 + \|\mathbf W\|_F^2 \leq \Ptd_E.
\end{cases}
\end{align}

\begin{lemma}\label{lemma:lm1}
The optimal solution to problem \eqref{eq:rDMax2} is
\begin{align}
\mathbf W^\star=\sqrt{\mu^\star} \tdh_{ED}\tdh_{SE}^H, \label{eq:Wstar}
\end{align}
where $\mu^\star$ is the optimal solution to the following optimization problem
\begin{align}
\gammatd^{\max}_D(\boldsymbol \rho)\triangleq
\begin{cases}\label{eq:rDMax3}
 \underset{\mu}{\max } &\  \gammatd'_D(\mu)\triangleq \frac{\left( |h_{SD}|+ \sqrt{\mu} \|\hat{\mathbf h}_{ED}\|\|\hat{\mathbf h}_{SE}\| \right)^2 \Ptd_S}{1+\mu \|\hat{\mathbf h}_{ED}\|^2} \\
\text{s.t.} & 0\leq \mu \leq \frac{\Ptd_E}{\Ptd_S\|\hat{\mathbf h}_{SE} \|^2 + 1}.
\end{cases}
\end{align}
\end{lemma}
\begin{IEEEproof}
We show Lemma~\ref{lemma:lm1} by construction. Suppose that an optimal solution to \eqref{eq:rDMax2} is given by $\mathbf W'=\sqrt{\mu'} \tilde{\mathbf W}$, with $\|\tilde{\mathbf W}\|_F=1$, and the resulting optimal value  is  $\gammatd'_D(\mathbf W')$. We then construct an alternative solution $\mathbf W''$ in the form of \eqref{eq:Wstar}, i.e., $\mathbf W''=\sqrt{\mu''} \tilde{{\mathbf h}}_{ED}\tilde{{\mathbf h}}_{SE}^H$, with
\begin{align}
\mu''=\mu' \frac{ \big |\hat{\mathbf h}_{ED}^H \tilde{\mathbf W} \hat{\mathbf h}_{SE}\big|^2}{\|\hat{\mathbf h}_{ED} \|^2 \|\hat{\mathbf h}_{SE} \|^2} \leq \mu',
\end{align}
where the last inequality follows from the sub-multiplicativity property of the Frobenius norm, i.e., $\|\mathbf A \mathbf B\|_F\leq \|\mathbf A\|_F \|\mathbf B\|_F$ for any conformable matrices (vectors) $\mathbf A$ and $\mathbf B$ \cite{361}.  We aim to show that the newly constructed matrix $\mathbf W''$ is feasible to problem \eqref{eq:rDMax2}, and also returns an objective value no smaller than $\gammatd'_D(\mathbf W')$. To this end, we first show the following:
\begin{align}
\Ptd_S\|\mathbf W'' \hat{\mathbf h}_{SE} \|^2 + \|\mathbf W''\|_F^2
&=\Ptd_S \mu' \frac{\big |\hat{\mathbf h}_{ED}^H \tilde{\mathbf W} \hat{\mathbf h}_{SE}\big|^2}{\|\hat{\mathbf h}_{ED}\|^2}+ \mu'' \\
& \leq \Ptd_S\mu' \|\tilde{\mathbf W} \hat{\mathbf h}_{SE}\|^2+ \mu' \label{eq:ineq1} \\
&\leq \Ptd_E, \label{eq:ineq2}
\end{align}
where \eqref{eq:ineq1} follows from $\mu''\leq \mu'$ and the Cauchy-Schwarz inequality $|\mathbf a^H \mathbf b|^2\leq  \|\mathbf a\|^2 \|\mathbf b\|^2$, $\forall \mathbf a, \mathbf b$; and  \eqref{eq:ineq2} is true since $\mathbf W'=\sqrt{\mu'}\tilde{\mathbf W}$ must satisfy the power constraint of problem \eqref{eq:rDMax2}. The above result shows that the newly constructed matrix $\mathbf W''$ also satisfies the power constraint at {\bf E}, and hence is feasible to problem \eqref{eq:rDMax2}. Furthermore, the following results can be obtained,
\begin{align}
\big |\hat{\mathbf h}_{ED}^H \mathbf W'' \hat{\mathbf h}_{SE}\big|& =\sqrt{\mu''}\|\hat{\mathbf h}_{ED}\|\|\hat{\mathbf h}_{SE}\|
=\sqrt{\mu'} \big |\hat{\mathbf h}_{ED}^H \tilde{\mathbf W} \hat{\mathbf h}_{SE}\big|\notag \\
&=\big |\hat{\mathbf h}_{ED}^H \mathbf W' \hat{\mathbf h}_{SE}\big|, \label{eq:numEq}\\
 \|\hat{\mathbf h}_{ED}^H \mathbf W''\|& = \sqrt{\mu''} \|\hat{\mathbf h}_{ED}\|
=\sqrt{\mu'} \frac{\big |\hat{\mathbf h}_{ED}^H \tilde{\mathbf W} \hat{\mathbf h}_{SE}\big|}{\|\hat{\mathbf h}_{SE}\|} \notag \\
& \leq  \sqrt{\mu'} \|\hat{\mathbf h}_{ED}^H \tilde{\mathbf W}\|=\|\hat{\mathbf h}_{ED}^H \mathbf W'\|.\label{eq:denLess}
\end{align}
Based on \eqref{eq:numEq} and \eqref{eq:denLess}, it is not difficult to conclude that $\gammatd'_D(\mathbf W'')\geq \gammatd'_D(\mathbf W')$. In summary, for any optimal solution $\mathbf W'$ to problem \eqref{eq:rDMax2}, we can always construct a feasible solution $\mathbf W''$ in the form of \eqref{eq:Wstar} that achieves no smaller objective value; thus, $\mathbf W''$ must also be optimal. Furthermore, problem \eqref{eq:rDMax3} is resulted by substituting \eqref{eq:Wstar} into \eqref{eq:rDMax2}. This completes the proof of Lemma~\ref{lemma:lm1}.
\end{IEEEproof}

The uni-variate optimization problem \eqref{eq:rDMax3} can be then solved by examining its first-order derivative. Together with Lemma~\ref{lemma:lm1}, the results in Theorem~\ref{theo:theo1} can be obtained. This completes the proof of Theorem~\ref{theo:theo1}.

\section{Proof of Theorem~\ref{theo:theo2}}\label{A:theo2}
To show Theorem~\ref{theo:theo2}, we first apply the triangle inequality to \eqref{eq:gammatdD2}, which yields
\begin{align}
\tilde{\gamma}_D(\boldsymbol \rho, \mathbf W)\geq \frac{\left( \big|h_{SD}\big|- \big| \hat{\mathbf h}_{ED}^H \mathbf W\hath_{SE}\big| \right)^2 P_S}{(1+\|\hat{\mathbf h}_{ED}^H \mathbf W\|^2)\sigma^2}, \ \forall \mathbf W,
\end{align}
where equality holds if and only if $\mathbf W$ is chosen such that the two effective signal paths from {\bf S} to {\bf D} add destructively, i.e., $\angle h_{SD} = \pi+\angle \hat{\mathbf h}_{ED}^H \mathbf W\hath_{SE}$. Furthermore, since an arbitrary phase shifting of $\mathbf W$ does not alter the feasibility of problem \eqref{eq:rDMin}, the optimal solution to \eqref{eq:rDMin} can be obtained by solving
\begin{align}
\gammatd^{\min}_D(\boldsymbol \rho)\triangleq
\begin{cases}\label{eq:rDMin2}
 \underset{\mathbf W}{\min } &\  \gammatd''_D(\mathbf W)\triangleq \frac{\big( |h_{SD}|- \big |\hat{\mathbf h}_{ED}^H \mathbf W \hat{\mathbf h}_{SE}\big| \big)^2 \Ptd_S}{1+\|\hat{\mathbf h}_{ED}^H \mathbf W\|^2} \\
\text{s.t.} & \Ptd_S\|\mathbf W \hat{\mathbf h}_{SE} \|^2 + \|\mathbf W\|_F^2 \leq \Ptd_E.
\end{cases}
\end{align}

Next, we derive the optimal structure of the  solution to \eqref{eq:rDMin2}. Recall that $\mathbf W$ is a matrix of dimension $r_0\times M$. Define an $r_0$-dimensional unitary matrix $\mathbf U\triangleq \left[\tdh_{ED} \ \mathbf U_{\perp} \right]$, where $\mathbf U_{\perp}\in \mathbb{C}^{r_0\times (r_0-1)}$ is the orthogonal complement of $\tdh_{ED}$ such that $\mathbf U^H \mathbf U=\mathbf I_{r_0}$. Similarly, define an $M$-dimensional unitary matrix $\mathbf V\triangleq \left[\tdh_{SE} \ \mathbf V_{\perp} \right]$, where $\mathbf V^H \mathbf V=\mathbf I_{M}$. Then any matrix $\mathbf W\in \mathbb{C}^{r_0\times M}$ can be expressed as
\begin{align}\label{eq:WMin}
\mathbf W=\mathbf U \mathbf Q \mathbf V^H=
\left[\tdh_{ED} \ \mathbf U_{\perp} \right]\left[\begin{matrix} q_{11} & \mathbf q_{12}^H \\ \mathbf q_{21} & \mathbf Q_{22} \end{matrix} \right] \left[ \begin{matrix} \tdh_{SE}^H \\  \mathbf V_{\perp}^H  \end{matrix} \right],
\end{align}
where $q_{11}\in \mathbb{C}$, $\mathbf q_{12}\in \mathbb{C}^{(M-1)\times 1}$, $\mathbf q_{21}\in \mathbb{C}^{(r_0-1)\times 1}$, and $\mathbf Q_{22}\in \mathbb{C}^{(r_0-1)\times (M-1)}$ are the new optimization variables after unitary transformations by $\mathbf U$ and $\mathbf V$. By substituting $\mathbf W$ with \eqref{eq:WMin}, we have the following results:
\begin{align}
&\big|\hat{\mathbf h}_{ED}^H \mathbf W \hat{\mathbf h}_{SE}\big|=|q_{11}| \|\hath_{ED}\|\|\hath_{SE}\| \label{eq:eq1}\\
&\|\tdh_{ED}^H \mathbf W\|^2 =\|\hath_{ED}\|^2 \left(|q_{11}|^2+ \|\mathbf q_{12}\|^2 \right) \label{eq:eq2} \\
&\|\mathbf W\hath_{SE}\|^2=\|\hath_{SE}\|^2 \left(|q_{11}|^2+ \|\mathbf q_{21}\|^2 \right) \label{eq:eq3}\\
&\|\mathbf W\|_F^2=|q_{11}|^2+ \|\mathbf q_{21}\|^2 + \|\mathbf q_{12}\|^2 + \|\mathbf Q_{22}\|_F^2. \label{eq:eq4}
\end{align}
It is observed from \eqref{eq:eq1}-\eqref{eq:eq4} that the objective value of problem \eqref{eq:rDMin2} is independent of $\mathbf q_{21}$ and $\mathbf Q_{22}$. Besides, the left hand side (LHS) of the power constraint in problem \eqref{eq:rDMin2} increases with $\|\mathbf q_{21}\|^2$ and $\|\mathbf Q_{22}\|_F^2$. Thus, without loss of optimality, we can set $\mathbf q_{21}=\mathbf 0$ and $\mathbf Q_{22}=\mathbf 0$. Furthermore, as both the objective value and the transmit power depend on $\mathbf q_{12}$ via its norm $\|\mathbf q_{12}\|$ only, we can assume without loss of optimality that $\mathbf q_{12}=\sqrt{z_2}\left[1\ 0 \cdots 0 \right]^T$ for $z_2\geq 0$. Similarly, we may assume $q_{11}=\sqrt{z_1}$ for $z_1\geq 0$ without loss of optimality. Therefore, it follows from  \eqref{eq:WMin} that the optimal solution to \eqref{eq:rDMin2} can be expressed as
\begin{align}
\mathbf W=\tdh_{ED}\left(\sqrt{z_1}\tdh_{SE} + \sqrt{z_2}\tdh_{SE}^\perp\right)^H.\label{eq:WMin2}
\end{align}
By substituting  \eqref{eq:WMin2} into problem \eqref{eq:rDMin2}, we obtain the optimization problem \eqref{eq:rDMin3} for determining the optimal weighting coefficients $z_1$ and $z_2$.

This completes the proof of Theorem~\ref{theo:theo2}.

\section{Proof of Theorem~\ref{theo:theo3}}\label{A:theo3}
Note that the objective value of problem \eqref{eq:rDMin3} is always non-negative, and it equals to zero if $z_1=|h_{SD}|^2/\big(\|\hath_{ED}\|^2\|\hath_{SE}\|^2\big)\triangleq z_1'$. Thus, if $z_1'$ is achievable, i.e., $z_1'\leq \tilde{P}_E/(1+\tilde{P}_S\|\hath_{SE}\|^2)$, the pair $(z_1', 0)$ is obviously the optimal solution to  \eqref{eq:rDMin3}. This corresponds to the first case of \eqref{eq:z1z2Opt}. For the remaining cases, it can be verified that at the optimal solution, the power constraint of \eqref{eq:rDMin3} should be satisfied with equality, since otherwise, one can always increase $z_2$ to further minimize the objective value. Therefore, the variable $z_2$ can be eliminated by substituting with $z_1$, and the problem reduces an uni-variate optimization problem, which can be solved by examining its first-order derivative. The details are omitted for brevity.

\section{Proof of Theorem~\ref{theo:thoe4}}\label{A:thoe4}
The key for proving Theorem~\ref{theo:thoe4} is to use the fact that all the three functions $\gammatd_E(\boldsymbol \rho)$, $\gammatd_D^{\max}(\boldsymbol \rho)$, and  $\gammatd_D^{\min}(\boldsymbol \rho)$ depend on $\boldsymbol \rho$ only via the term $\|\hath_{SE}(\boldsymbol \rho)\|^2=\mathbf h_{SE}^H \diag(\boldsymbol \rho)\mathbf h_{SE}=\sum_{m=1}^M \rho_m |h_{SE,m}|^2$, where $h_{SE,m}$ denotes the $m$th element of $\mathbf h_{SE}$. Assume that $(\boldsymbol \rho', \gammatd_D')$ is an optimal solution to (P3), with the $m$th element of $\boldsymbol \rho'$ given by $0\leq \rho_m'\leq 1$, $m=1,\cdots, M$. We construct a new power splitting vector $\boldsymbol \rho''=\rho'' \boldsymbol 1$, with $\rho''=\left(\sum_{m=1}^M \rho_m' |h_{SE,m}|^2\right)/\left(\sum_{m=1}^M |h_{SE,m}|^2\right)$. It is obvious that $0\leq \rho'' \leq 1$, and thus the constraint $\boldsymbol 0 \preceq \boldsymbol \rho'' \preceq \boldsymbol 1$ is satisfied. Furthermore, we also have $\|\hath_{SE}(\boldsymbol \rho'')\|^2=\|\hath_{SE}(\boldsymbol \rho')\|^2$, and hence $\gammatd_E(\boldsymbol \rho'')=\gammatd_E(\boldsymbol \rho')$, $\gammatd_D^{\max}(\boldsymbol \rho'')=\gammatd_D^{\max}(\boldsymbol \rho')$, and  $\gammatd_D^{\min}(\boldsymbol \rho'')=\gammatd_D^{\min}(\boldsymbol \rho')$. Therefore, the pair $(\boldsymbol \rho'', \gammatd_D')$ is also an optimal solution to (P3). This completes the proof of Theorem~\ref{theo:thoe4}.

\section{Proof of Theorem~\ref{theo:theo5}}\label{A:theo5}
The proof of Theorem~\ref{theo:theo5} is similar to that of Theorem~\ref{theo:thoe4}, which exploits the fact that the power splitting vector $\boldsymbol \rho$ affects the SNRs at both {\bf D} and {\bf E} only via the term $\|\hath_{SE}(\boldsymbol \rho)\|^2=\sum_{m=1}^M \rho_m |h_{SE,m}|^2$. Let $\boldsymbol \rho_{\mathrm{ups}}^\star=\rho_{\mathrm{ups}}^{\star} \mathbf 1$  be the optimal UPS vector to problem (P3). If $\rho_{\mathrm{ups}}^\star=0$ or $\rho_{\mathrm{ups}}^\star=1$, then Theorem~\ref{theo:theo5} is already satisfied. Thus, we assume $0<\rho_{\mathrm{ups}}^{\star} <1$. In this case, it can be verified that there always exists an integer $m'\in \{1,\cdots, M\}$ such that both the following inequalities hold,
\begin{align}
&\sum_{m=1}^{m'-1} \big|h_{SE,[m]}\big|^2 < \rho_{\mathrm{ups}}^{\star} \sum_{m=1}^M \big|h_{SE,m}\big|^2, \\
& \sum_{m=1}^{m'} \big|h_{SE,[m]}\big|^2 \geq \rho_{\mathrm{ups}}^{\star}  \sum_{m=1}^M \big|h_{SE,m}\big|^2,
\end{align}
where $[\cdot]$ is the permutation operation such that $|h_{SE,[1]}|^2\geq \cdots \geq |h_{SE,[M]}|^2$. As a result, we define a new power splitting vector $\boldsymbol \rho_{\mathrm{bps}}$ with binary power splitting (BPS) over $M-1$ receiving antennas such that
\begin{align}\label{eq:rhoBps}
\rho_{\mathrm{bps},m} =
\begin{cases}
1, &  \ m=[1],\cdots, [m'-1], \\
\rho_{\mathrm{bps}}, & \ m=[m'], \\
0, & m=[m'+1], \cdots, [M],
\end{cases}
\end{align}
where
\begin{align}
\rho_{\mathrm{bps}}=\frac{\rho_{\mathrm{ups}}^{\star} \sum_{m=1}^M |h_{SE,m}|^2-\sum_{m=1}^{m'-1} |h_{SE,[m]}|^2}{|h_{SE,[m']}|^2}.
\end{align}
It can be verified that $0<\rho_{\mathrm{bps}}\leq 1$, and hence  $\mathbf 0 \preceq \boldsymbol \rho_{\mathrm{bps}} \preceq \mathbf 1$ is satisfied. Furthermore, we have  $\|\hath_{SE}(\rho_{\mathrm{bps}} )\|^2=\|\hath_{SE}(\rho_{\mathrm{ups\star}})\|^2$. Thus, $\boldsymbol \rho_{\mathrm{bps}} $ must also be an optimal solution to problem $\mathrm{(P3)}$. This completes the proof of Theorem~\ref{theo:theo5}.

\balance
\bibliographystyle{IEEEbib}
\bibliography{IEEEabrv,IEEEfull}

\end{document}